\documentclass[fleqn,usenatbib]{mnras}

\usepackage{newtxtext,newtxmath}

\usepackage[T1]{fontenc}

\DeclareRobustCommand{\VAN}[3]{#2}
\let\VANthebibliography\thebibliography
\def\thebibliography{\DeclareRobustCommand{\VAN}[3]{##3}\VANthebibliography}

\usepackage{graphicx}	
\usepackage{amsmath}	
\usepackage{multirow}
\usepackage{subcaption}

\title[Heating due to Cosmic Rays]{Impact of cosmic rays on the global 21-cm signal during cosmic dawn}

\author[A. Bera et al.]{
Ankita Bera,$^{1}$\thanks{E-mail: ankita1.rs@presiuniv.ac.in}
Saumyadip Samui,$^{1,3}$
Kanan K. Datta$^{1,2}$
\\
$^{1}$Department of Physics, Presidency University, 86/1 College Street, Kolkata 700073, India
\\
$^{2}$Department of Physics, Jadavpur University, 188, Raja S.C. Mallick Rd, Kolkata 700032, India \\
$^{3}$School of Astrophysics, Presidency University, 86/1 College Street, Kolkata 700073, India
}

\date{Accepted XXX. Received YYY; in original form ZZZ}

\pubyear{2021}

\begin{document}
\label{firstpage}
\pagerange{\pageref{firstpage}--\pageref{lastpage}}
\maketitle

\begin{abstract}
It is extremely important to understand the processes through which the thermal state of the inter-galactic medium (IGM) evolved in the early universe in order to study the evolution of HI 21-cm signal during cosmic dawn. 
Here we consider the heating of the IGM due to cosmic ray protons generated by the supernovae from both early Pop~III and Pop~II stars. The low energy cosmic ray protons from Pop~III supernovae can escape from minihalos and heat the IGM via collision and ionization of hydrogen. Furthermore, high energy protons generated in Pop~II supernovae can escape the hosting halos and heat the IGM via magnetosonic Alfv\'en waves. We show that the heating due to these cosmic ray particles can significantly impact the IGM temperature and hence the global 21-cm signal at $z\sim 14-18$. The depth, location, and duration of the 21-cm absorption profile are highly dependent on the efficiencies of cosmic ray heating. In particular, the EDGES signal can be well fitted by the cosmic ray heating along with the Lyman-$\alpha$ coupling and the dark matter-baryon interaction that we consider to achieve a `colder IGM background'. Further, we argue that the properties of cosmic rays and the nature of first generation of stars could  be constrained by accurately measuring the global 21-cm absorption signal during the cosmic dawn.
\end{abstract}

\begin{keywords}
galaxies: high-redshift -- dark ages, reionization, first stars -- cosmic rays -- dark matter -- cosmology: theory 
\end{keywords}



\section{Introduction}
{The globally averaged 21-cm signal from neutral hydrogen (HI) is an important tool for studying early epochs such as cosmic dawn (CD) and the epoch of reionization (EoR).}
{The first generation of stars, i.e., Population~III (Pop~III) stars
are likely to impact the evolution of the HI 21-cm signal during the cosmic dawn \citep{Mirocha2018, Mebane_2020, Munoz2022}.} Unlike today's metal-enriched stars, the Pop~III stars are believed to have formed in metal-free gas clouds through molecular hydrogen (H$_2$) cooling mechanism in minihalos \citep{bromm99, abel02}.
{Although there has been a lot of work to study the formation and properties of these Pop~III stars through different techniques such as numerical simulations \citep{ Wise_2007, Stacy_2012, Hirano_2015, Xu_2016, Susa2019, Sugimura2020, Wollenberg2020}, analytical argument \citep{Mckee_2008} and semi-analytical modelling \citep{Trenti_2009, Visbal_2018}, the direct observation of these high redshift stars is yet to be done and may be challenging even for the upcoming missions. However properties of Pop~III stars can be studied and constrained through their effects on the global 21-cm signal \citep{madau18, Mirocha_2019, atri20, Mebane_2020, Hibbard2022, Magg2021, Gessey2022}.}

Indeed, observations of global HI 21-cm absorption signal by experiments such as the Experiment to Detect the Epoch of Reionization Signature (EDGES) \citep{EDGES18} have opened up the possibilities to study the evolution of the early stars and galaxies and the thermal state of the inter-galactic medium (IGM) during cosmic dawn. The detected signal has an absorption depth of $0.5^{+0.5}_{-0.2}$~K centred at frequency $78\pm 1$~MHz or redshift $z \sim 17$. The `U' shaped 21-cm signal caries signatures of early Lyman-$\alpha$ (Ly-$\alpha$) coupling and heating of the IGM in the redshift range of $\sim 14-22$. {However, there seems to be quite controversity exists with the detected EDGES signal. For example, recent observations by Shaped Antenna measurement of the background RAdio Spectrum (SARAS~3) \citep{Saurabh_2021} has claimed that the EDGES profile may not be of astrophysical origin.} {There are also concerns that the unusual signal may arise due to unaccounted systematics \citep{Hills_2018, Bradley2019, Singh2019, Sims2020}. Nonetheless, several explanations have been proposed in order to explain the unusual absorption trough as the absorption depth of the detected signal is almost two times larger than the strongest prediction  \citep[see e.g.,][]{Barkana18Nature, Munoz18, Fraser_2018, Pospelov_2018, Slatyer2018, Feng_2018, Ewall_2018, Mirocha_2019, Fialkov_2019, Mebane_2020, Ewall2020}.}

The 21~cm differential brightness temperature depends on the background radio signal, the hydrogen spin temperature and the kinetic temperature of the IGM. The depth of the detected signal at $z\sim 17$ could possibly be explained either by a colder IGM achieved via cold dark matter and baryon interaction {\citep{Barkana18Nature, Munoz18} } or by an excess radio background over the cosmic microwave background radiation (CMBR) \citep{Fraser_2018, Ewall_2018, Fialkov_2019}. However the rise of the absorption signal in the redshift range $z\sim 16-14$ requires the IGM temperature to increase very rapidly to match with the background radio temperature. {The X-ray heating by the first generation of galaxies/black holes is one such possibility which has been widely studied \citep{pritchard2007, baek09, mesinger13, ghara14, Pacucci2014, Fialkov2014, fialkov14b, Arpan2017, Ma2021}. Other possibilities such as heating due to the primordial magnetic field \citep{SS05, Bera_2020}, shocks \citep{xu21}, Ly-$\alpha$ photons \citep{Madau1997, Chuzhoy2006, ghara2020, Ciardi2010, Reis2021, mittal21}, CMB photons \citep{Venumadhav2018}  have also been explored.} However, all these mechanisms have their own parameter spaces that are poorly constrained in those high redshift universe. Thus, how and when did the IGM heating take place during cosmic dawn still remains an unsettled issue.   
 
In this work, we explore another possible source of IGM heating, namely, the cosmic rays (CR) generated from the first generation of galaxies and show that they play an important role in the IGM heating and thus on the 21~cm signal. The possibilities of heating and ionization of the IGM by cosmic rays from young galaxies are already discussed by \citet{Ginzburg_1966, Nath_Biermann_1993, Samui_2005, Samui_2018} in the context of reionization  and post-reionization era. Further, impact of cosmic rays generated from microquasars on the reionization was discussed in \citet{Tueros_2014}. However there has been only a handful of earlier works which discuss the effect of cosmic ray heating during the cosmic dawn and its consequences on the global 21-cm signal \citep{Sazonov_2015, Leite_2017, Jana_2019}. {Among them, \citet{Sazonov_2015} has adopted a simplified approach by assuming that a small fraction of supernovae kinetic energy goes into the IGM heating without any detailed modeling of the energy transfer. They did not consider contribution from Pop~III and Pop~II stars separately as well. Moreover, the effects of cosmic rays coming from early epochs are also not considered by \citet{Sazonov_2015} as well as in \citet{Jana_2019}. On the other hand, \citet{Leite_2017} did modeled the propagation of cosmic rays in the IGM but neither considered the detailed model of the star formation in Pop~III and Pop~II galaxies nor the 21~cm signal.}

{
Here we investigate, in more details, the heating of the IGM by cosmic rays from Pop~II and Pop~III stars and their impact on the global HI 21~cm signal during cosmic dawn. We model contributions of cosmic rays from Pop~III stars and Pop~II stars separately as they are expected to produce different supernovae. Further, we consider more detailed modelling of star formation by taking into account various feedbacks like Lyman-Warner feedback, supernova and radiative feedback etc. Moreover, our model accounts for the evolution/propagation of cosmic ray particles from previous redshifts and the energy deposition by these particles are computed in detail instead of adopting a simplified approach  as followed in previous studies. Our study also focuses on the standard IGM scenario along with a scenario where dark matter-baryon interaction is considered  in light of recent observations of global 21~cm signal. Dark matter-baryon interaction makes the IGM colder as compared to the IGM in the standard scenario and we refer this as `cold IGM' scenario. Finally, we show that our cosmic ray heating  model can explain the EDGES observations with reasonable choice of model parameter and thus establishing the importance of detailed modeling of cosmic ray heating. The main purpose of this work is to investigate  cosmic rays from Pop III and Pop II stars as a potential source of IGM heating during cosmic dawn.  In addition, we compared the efficiency of cosmic rays heating with the more conventional IGM heating mechanism  by X-rays during cosmic dawn.}

The paper structure is as follows. In Section~\ref{main_sec:formalism} we outline the basics of global 21-cm signal and our semi analytical models for Pop~III and Pop~II star formation including Lyman-Werner feedback. We also describe there the evolution of cosmic rays, Lyman-$\alpha$ coupling and the interaction mechanism between dark matter and baryons. 
We present our results of the star formation, the kinetic temperature, spin temperature and finally the differential brightness temperature highlighting the effect of cosmic ray heating in Section~\ref{main_sec:results}. This section also highlights the changes in differential temperature due to the variation in assumed model parameters.
Finally in Section~\ref{main_sec:summary} we present our conclusions including the summary of our results.
We assume a flat $\Lambda$CDM cosmology throughout this paper with the cosmological parameters obtained from recent Planck 2018 \citep{Planck18} observation, i.e., $\Omega_{\Lambda} = 0.69$, $\Omega_{\rm m} = 0.31$, $\Omega_{\rm b} = 0.049$, and the Hubble parameter $H_0 = 67.66$~km/s/Mpc.



\section{Formalism}
\label{main_sec:formalism}
\subsection{Global 21-cm signal}
\label{sec:21cm_th}
We begin with the 21-cm radio signal as received from the early universe due to the presence of neutral hydrogen.
The globally averaged differential brightness temperature ($T_{21}$) is the measure of the redshifted 21-cm signal at redshift $z$. This can be written as\citep{pritchard12},
\begin{equation}
    T_{21} \approx 27\times10^{-3} x_{\rm HI} \left( 1-\frac{T_{\gamma}}{T_s} \right) \left(\frac{\Omega_{\rm b} h^2}{0.02} \right) \left(\frac{0.15}{\Omega_{\rm m} h^2} \right)^{0.5} \left( \frac{1+z}{10} \right)^{0.5} \, {\rm K} .
    \label{eq:T21}
\end{equation}
Here $T_{\gamma}$ is the temperature of the background radio emission which is dominated by the CMBR and $T_s$ is the hydrogen spin temperature which is determined by the relative population of the two hydrogen ground state hyperfine levels. Further, $x_{\rm HI}$ is the neutral hydrogen fraction which is determined by the physical processes such as recombination, photoionization and collisional ionization. The ionization fraction $x_e$ ($=1-x_{\rm HI}$) can be calculated as \citep{Peebles1968},
\begin{multline}
    \frac{dx_e}{dz} = \frac{1}{H(z)\,(1+z)} \biggl[ C_p \left( \alpha_e\,x_e^2 n_{\rm H} -\beta_e\,(1-x_e) \, e^{-\dfrac{h_p\nu_\alpha}{k_B T_g}} \right) \\ 
    - \gamma_e\,n_H (1-x_e)x_e \biggr].
    \label{eq:ion_frac}
\end{multline}
Here $C_p$ is the Peeble's factor, {$h_p$ is the Planck's constant}, $\alpha_e$ is the recombination co-efficient, $\beta_e$ is the photoionization coefficient as given in \citet{Seager1999,Seager2000} and $\gamma_e$ is the collisional ionization coefficient \citep{Minoda17}.
{Note that Eq.~\ref{eq:ion_frac} holds for high redshift universe only.
In the late time the ionizing photons and cosmic rays from the first generation of galaxies are likely to alter the ionization fraction $x_e$. However, we have checked that the cosmic rays can change the ionization fraction at most $10^{-3}$ for the supernova kinetic energy we considered here which is similar to \citet{Sazonov_2015}. Similar results were also obtained by \citet{Samui_2005}. Since we are interested in 21~cm signal in the redshift range 10 to 20 where the 21~cm signal is mostly governed by the temperature differences between the IGM and the background radio signal rather than the ionization fraction of the hydrogen which is at most 0.1 by redshift $z=10$ thus only altering the 21 cm signal at most 10\% level \citep{furlanetto2006}. One should consider the contribution of UV photons from the first stars in order to model the ionization fraction more accurately that we are not taking into account in this particular work as we are interested in the cosmic dawn.}

The spin temperature ($T_s$) in Eq.~\ref{eq:T21} is governed by three coupling mechanisms: (i) radiative transition due to the absorption and stimulated emission of CMB photons (couples $T_s$ and CMBR temperature $T_{\gamma}$, (ii) spin flip transition due to atomic collisions (couples $T_s$ and gas kinetic temperature $T_g$) and (iii) the Wouthuysen-Field effect \citep{Wouthuysen_1952, Field_1959} which also couples $T_s$ and $T_g$.
Hence to determine $T_{21}$ we need to know the gas kinetic temperature $T_g$ along with the ionization fraction. The evolution of $T_g$ as a function of redshift can be obtained by solving the equation  \citep{Kompaneets_1957,Peebles_1993, Seager1999}, 
\begin{multline}
    \frac{dT_g}{dz} = \frac{2T_g}{1+z} - \frac{8\sigma_T a_{\rm SB} T_{\gamma}^4}{3m_e c H(z) (1+z)}\left(T_\gamma-T_g\right) \frac{x_e}{1+x_e}  \\
    - \frac{2}{3k_B} \sum_i Q_i .
	\label{eq:Tg}
\end{multline}
The first two terms on the R.H.S. arise due to adiabatic cooling of baryonic gas due to  expansion of the universe and the Compton heating due to the interaction between CMBR and free electrons, respectively. Further, $ k_{\rm B}$, $ \sigma_{\rm T}$ {are the Boltzmann constant, Thomson scattering cross-section respectively and $a_{\rm SB}=4 \sigma_{\rm SB}/c$, where $\sigma_{\rm SB}$ is the Stefan Boltzmann constant}. The third term $Q_i$ includes other heating and cooling mechanisms such as heating due to cosmic rays and cooling/heating due to the possible interaction between dark matter and baryons. These processes are described in the subsequent sections.

Note that, both the cosmic rays and Ly-$\alpha$ photons are generated by early galaxies. Their production depends on the star formation mechanisms which we model in the following section.


\subsection{Star formation in early universe}
\label{sec:SFRD_th}

In the hierarchical structure formation scenario, galaxies are formed inside dark matter halos and the halo mass function can provide the number density of such halos at different redshifts.
The differential halo mass function or, in short, halo mass function is the number of halos in the mass range  of $M$ and $M$+$dM$ per unit comoving volume. This can be written as \citep[see][for a review]{Reed_2007}, 
\begin{equation}
    \frac{dn(M, z)}{dM} dM= \frac{\bar \rho_0}{M} f(\sigma) \left| \frac{d \ln \sigma}{dM} \right|dM.
\end{equation}
Here $\bar \rho_0$ is the total matter density of the universe at present
and $\sigma(M,z) $ is the RMS fluctuations in the {dark matter} density field at scales corresponding to mass $M$ and redshift $z$.
Note that, the function $f(\sigma)$ depends on the particular form of the halo mass function that one adopts. There are several prescriptions of the halo mass function, and among them Press–Schechter formalism \citep{press1974} and Sheth-Tormen formalism \citep{S-T_1999} are widely used.
In this work, we adopt the Sheth-Tormen mass function as it fits well with the mass functions obtained from numerical simulations for a broader range of masses and redshifts. Once the halo collapsed,
the baryonic gas inside the dark matter halos would form stars if the gas cools. 

The comoving global star formation rate densities can be calculated as \citep{Samui_2007},
\begin{equation}
    {\rm SFRD}(z) = \int_{z}^{\infty} dz_c \int_{M_{\rm min}}^{\infty} dM^{\prime} \dot M_{*}(M^{\prime},z,z_c) {\frac{d^2n(M^{\prime},z_c)} {dz_c dM^{\prime}}}, 
    \label{eq:SFR_th}
\end{equation}
where $z_c$ is the collapse redshift of the dark matter halos which is forming stars at a rate $\dot M_{*}$ at redshift $z<z_c$. 
Here, we assume that the redshift derivative of $\frac{dn(M, z)}{dM}$ in equation~\ref{eq:SFR_th} provides the formation rate of dark matter halos \citep{Samui_2009}. Further, $M_{\rm min}$ is the lower mass cut-off of halos that can form stars. Note that, $\dot M_{*}$ depends on the halo mass $M$ and the cooling mechanisms.
Our models of Population~III (Pop~III) and Population~II (Pop~II) stars are described below in detail.


\subsubsection{Pop~III star formation}
\label{subsec:PopIII}
In this section we describe the Pop~III star formation model in  small mass halos.
We mostly follow \citet{Mebane_2018} for this purpose. 
Pop~III stars are the first generation metal free stars that form inside the minihalos where the gas can cool by the H$_2$ cooling only.
The star formation mechanism in these halos are not fully understood and there is no consensus about the initial mass function of Pop~III stars yet \citep{abe2021,parsons2021,lazar2022}. Pop~III stars could be very massive ($>100 M_{\odot}$) if the Jeans mass clump does not experience several fragmentation \citep{bromm99,abel02}. Due to the small size of their host halos these stars are likely to form in small numbers, possibly in isolation.
{Most of them are expected to be short-lived and either explode as a supernova or directly collapse into a black hole depending on their masses \citep{Mebane_2018}. Stars, having a mass in the range $40$~M$_{\odot}$ to $140$~M$_{\odot}$ and also above $260$~$\rm M_{\odot}$, are likely to collapse into a black hole. On the other hand, stars below $40$~M$_{\odot}$ preferably end their lives in a core-collapse supernova of energy $\sim 10^{51}$ erg. The intermediate mass range ($140$~$\rm M_{\odot}$ - $260$~$\rm M_{\odot}$) stars are likely to explode as a pair-instability supernova that release a kinetic energy of $\sim 10^{52}$ erg \citep{Wise2008, Greif2010}.}
{Here for the demonstration purpose, we assume that a Pop~III star of mass $145$~$\rm M_{\odot}$ is being formed per galaxy which is likely to explode as a SNe of energy $\sim 10^{52}$ erg \citep[mid mass range of ][]{Mebane_2018}. The variation of the IMF will be taken care by the parameters $\epsilon_{\rm III}$ described later. } 

Note that, the H$_2$ cooling depends on the amount of molecular hydrogen present in the halo.
{The molecular hydrogen forms through $\rm H^-$ and $e+$ catalysis. It can be destroyed by CMBR photons at high redshift but it get destroyed more by the presence of Lyman-Werner (LW) photons during the cosmic dawn that we model later in this section.}
Hence, the formation and destruction rates of $\rm H^-$ need to be balanced to get the desired H$_2$ in a halo.
\citet{Tegmark_1997} found that this fraction of molecular hydrogen varies with the halo's virial mass/temperature, and can be approximated as,
\begin{equation}
    f_{\rm H_2} \approx 3.5 \times 10^{-4} T_3^{1.52},
    \label{eq:fH2}
\end{equation}
where $T_3 = T_{\rm vir}/10^{3}$ K. Further note that, there is a critical threshold fraction of molecular hydrogen required for the cooling to be efficient to form Pop~III stars. This critical threshold fraction is given by \citep{Tegmark_1997}, 
\begin{equation}
    f_{\rm crit,H_2} \approx 1.6 \times 10^{-4} \left( \frac{1+z}{20} \right)^{-3/2} \left( 1+ \frac{10 T_3^{7/2}}{60+T_3^4} \right)^{-1} exp\left( \frac{0.512 K}{T_3} \right).
    \label{eq:f_cH2}
\end{equation}
The molecular cooling becomes efficient once $f_{\rm H_2} > f_{\rm crit,H_2}$ and this sets the minimum mass of halo that can host Pop~III stars (i.e. $M_{\rm min}$ in Eq.~\ref{eq:SFR_th}) in absence of any feedback.

{
The Pop~III stars are the first generation of metal free stars likely to form in the small mass haloes at high redshifts. The transition from Pop~III to Pop II star formation are governed by the amount of metal present in the star forming haloes. Several authors have modeled such transition with different model prescription. For example, \citet{sun_2021} have used a critical halo mass governed by the metal mixing time scale for the maximum halo mass that can host Pop~III stars and have shown such simple model can reproduce the Pop~III star formation reasonably well with detailed metal mixing model such as \citet{Mebane_2018}. They have also shown that the Pop~III star formation in two extreme models vary less than an order of magnitude. Motivated by that we assume here that the critical mass for transition from Pop~III to Pop~II stars occurs at a fixed mass which is the atomic cooling mass (corresponding to a virial temperature $T_{\rm vir}=10^4~$K). A detailed modelling may be more appropriate. However, as we will show that even with this simple assumption the resulting Pop~III star formation matches reasonable well with the detailed metal mixing models \citep[i.e.][]{Mebane_2018, sun_2021}.
}

As soon as the first generation of stars form they produce a background of Lyman-Werner radiation. The LW band consists of photons having energy in the range of {11.2}-13.6 eV which can photo-dissociate the molecular H$_2$. Thus, if a halo is present in a LW background, {it  reduces the concentration of H$_2$ to a level that H$_2$ cooling becomes inefficient and the halo cannot host any stars further.}
As a consequence, the minimum mass of a halo that can sustain the Pop~III star formation increases. Therefore, a self-consistent calculation is required to determine the minimum halo mass that can harbour Pop~III stars. 
Given the star formation, the background Lyman-Werner flux ($\rm J_{LW}$) can be calculated as \citep{visbal14},
\begin{equation}
    J_{\rm LW}(z) = \frac{c}{4 \pi} \int_z^{z_m} \frac{dt}{dz^{'}} (1+z)^3 \epsilon(z^{'}) dz^{'}.
    \label{eq:J_LW}
\end{equation}
Here $z_m$ is the maximum redshift that a LW photon can travel through IGM undisturbed and redshift into a Lyman series line, and can be written as, $\frac{1+z_m}{1+z} = 1.04$ \citep{visbal14}. $c$ is the speed of light and $\epsilon(z)$ is the specific LW comoving luminosity density which is given by,
\begin{equation}
    \epsilon(z) = \int_{z}^{\infty} dz_c \int_{M_{\rm min}}^\infty {\frac{d^2n(M,z_c)}{dz_c dM}} \frac{\dot M_{*}}{m_p} \left( \frac{N_{\rm LW} E_{\rm LW}}{\Delta \nu_{\rm LW}} \right) dM,
    \label{eq:eps_z}
\end{equation}
where $m_p$ is the proton mass, $E_{\rm LW} = 11.9\,{\rm eV}$ is the average energy of a LW photon and $\Delta \nu_{\rm LW} = 5.8 \times 10^{14}$ Hz \citep{Mebane_2018} is the LW frequency band.
Further, $N_{\rm LW}$ is the number of LW photons produced per baryon of stars. 
Note that, both Pop~III and Pop~II stars (that we discuss later) produce LW photons. For Pop~III stars we take $N_{\rm LW} = 4800$  and $N_{\rm LW} = 9690$ for Pop~II stars \citep{Barkana_2005, Pritchard_2006}. 

In the presence of the LW background
given above \citet{machacek01} showed that the mass of a
halo that can host Pop~III stars
must have virial temperature above a critical temperature given by,
\begin{equation}
    \frac{T_{\rm crit}}{1000 K} \sim 0.36 \left[ (\Omega_{\rm b} h^2)^{-1} (4 \pi J_{\rm LW}) \left( \frac{1+z}{20} \right)^{3/2} \right]^{0.22},
    \label{eq:LW_mass}
\end{equation}
where $J_{\rm LW}$ is in unit of $10^{-21} \rm{erg \, s^{-1}cm^{-2}Hz^{-1}sr^{-1}}$. We use this $T_{\rm crit}$ to calculate $M_{\rm min}$ in presence of LW background if that $M_{\rm min}$ is greater than the $M_{\rm min}$ obtained using the condition given in Eq.~\ref{eq:f_cH2}.


\subsubsection{Pop~II star formation}
\label{subsec:PopII}
In this section we outline the star formation model in atomic cooling halos with the virial temperature greater than $10^4~$K (Pop~II stars) that we adopt from \citet{Samui_2014}. We choose this SNe feedback regulated star formation model as it can successfully describe
the galaxy luminosity functions upto $z=10$, and can explain the observed stellar mass in galaxies of mass ranges from $10^{7} M_{\odot} \leq M \leq 10^{13} M_{\odot}$. {Note that, the model is well calibrated till $z=10$ and it is extrapolated to high redshift which may introduce some biases.} We briefly describe the model here.

In the presence of SNe feedback, the star formation rate of a galaxy of total baryonic mass $M_b$, at a time $t$ after the formation of dark matter halo, can be written as,
\begin{equation}
    \dot M_* = \frac{M_b f_* f_t}{\tau [f_t (1+\eta_\omega)-1]} \left[e^{-\frac{t}{\tau}} - e^{-f_t (1+\eta_\omega)\frac{t}{\tau}}\right].
\end{equation}
Here, the amount of SNe feedback in the form of galactic outflows has been characterised by the parameter $\eta_{\omega}$ which is defined as $\dot M_{\omega}=\eta_{\omega} \dot M_*$, $\dot M_{\omega}$ being the mass outflow rate.
It depends on the circular velocity ($v_c$) of the galaxy as well as
the driving mechanism of the outflow. We consider $\eta_{\omega}=(v_c/100~{\rm km/s})^{-2}$ that describes the outflows driven by the cosmic rays along with the hot gas produced by the SNe. The normalization constant is chosen to fit the UV luminosity functions of high redshift galaxies \citep[see][for details]{Samui_2014}. 
The total baryonic mass is related to the halo mass as $M_b=(\Omega_{\rm b}/\Omega_{\rm m}) M$. Further, the dimensionless parameter $f_t$ fixes the duration of the star formation activity in terms of the dynamical time, $\tau$, of the halo.
It is assumed that once the dark matter halo virialises and accretes the baryonic matter, a fraction, $f_*$ of the gas gets cooled and becomes available for star formation.
Hence, $f_*$ characterises the star formation efficiency and has been optimized to fit various available observation of high redshift galaxies.
{Further, note that, along with the SNe feedback, our Pop~II star formation model includes other feedbacks such as the radiative feedback from the ionizing photons, and AGN feedback as well. We are not describing all these feedback processes here in details, however the same can be found in \citet{Samui_2007}. As the ionization fraction is likely to be low at $z > 10$, the contribution from radiative feedback as well as AGN feedback are comparatively small at the redshift of our interest in present work. }


\subsection{Heating due to Cosmic rays}
\label{subsec:CR_th}

Cosmic rays are generated in the termination shock of the supernova explosions originated from both Pop~III and Pop~II stars. A significant fraction, $\epsilon \sim 0.15$ of the SNe kinetic energy ($E_{\rm SN}$) gets injected into the cosmic rays \citep{Hillas_2005, Caprioli_2014}. Thus the average rate of energy injection per unit physical volume into the cosmic rays (in units of ${\rm erg\,s^{-1} cm^{-3}}$) can be calculated as \citep{Samui_2005},
\begin{equation}
    \dot E_{\rm CR}(z) = 10^{-30} \epsilon \left( \frac{E_{\rm SN}}{10^{51} \, \rm erg} \right) f_{\rm SN} \left( \frac{{\rm SFRD}(z)}{\rm M_{\odot}\, yr^{-1} \, Mpc^{-3}} \right) (1+z)^3 .
    \label{eq:E_CR}
\end{equation}
Here the SFRD($z$) is obtained from Eq.~\ref{eq:SFR_th} for Pop~III and Pop~II stars. Further, $f_{\rm SN}$ is the number of SNe explosion per unit solar mass of star formation. {As already mentioned in the previous section, we assume a single star of 145 $M_{\odot}$ is formed in minihalos and explodes as a supernova having energy $10^{52}$ erg \citep[mid mass range of][]{Mebane_2018}.} Thus in our model, $f^{\rm SN} \approx 1/145$ and $E_{\rm SN} \sim 10^{52}$ erg for Pop~III stars.
{In case of Pop~II stars, we assume that a supernova of energy $E_{\rm SN} \sim 10^{51}$ erg is formed per 50 $M_{\odot}$ of a star having $1-100$ $M_{\odot}$ of Salpeter IMF and hence, $f_{\rm SN} = 0.02$ \citep{Samui_2005}.}

We note that, the cosmic ray protons mostly contribute to the heating of the IGM. In case of a supernova exploding in minihalos, low energy protons ($\leq$30 MeV) can escape the halo and heat the inter-galactic medium via collision with free $e^-$, and ionization of neutral hydrogen. These SNe are more energetic compared to the core-collapse SNe, and the shock front reaches the virial radius within the Sedov-Taylor (ST) phase itself. Thus the cosmic rays are generated outside the virial radius and get injected to the IGM easily \citep{Sazonov_2015}. This is contrary to  massive atomic cooling halos hosting Pop~II stars where the low energy protons get confined within the halo and only high energy protons can escape into the IGM and contribute to the heating \citep{Samui_2005}.
 
In general, the cosmic ray proton spectra can be modelled as a power law in the momentum space \citep{schlickeiser2002cosmic} which is given by,
\begin{equation}
    {\frac{dn_{\rm CR}(p,z)}{dt}} dp = \dot N_0(z) \left( \frac{p}{p_0} \right)^{-q} dp,
    \label{eq:p_spectra}
\end{equation}
where, $\dot N_0(z)$ (along with $p_0$) is the normalisation factor which is determined from the total available energies of cosmic rays. It is calculated by integrating $E(p) \dot N_0 (p/p_0)^{-q} dp$  with a low energy cut-off of $10$ keV and equating to $\dot E_{\rm CR}(z)$ (Eq.~\ref{eq:E_CR}). The slope of the spectrum, $q$ has a typical value of $2.2$ \citep{schlickeiser2002cosmic} which matches quite well with observations.

Unlike UV photons, the number density of cosmic ray protons in the IGM at a redshift, $z$ is contributed by the cosmic rays generated at $z$, and cosmic rays that are injected and evolved from higher redshift, $z_i>z$. 
Therefore, the physical number density of cosmic rays at a redshift $z$, having velocity between $\beta$ and $\beta + d\beta$ is given by,
\begin{equation}
    N_{\rm CR}(\beta, z, z_0) = \int_{z_0}^{z} dz_i \frac{dn(z_i, p_i)}{dz_i} \frac{dp_i}{d\beta_i} \frac{d\beta_i}{d\beta} \left( \frac{1+z}{1+z_i} \right)^3 .
    \label{eq:N_cr}
\end{equation}
Here, $z_0$ is the initial redshift of cosmic ray injection by the first generation of stars and we have taken it to be $z= 50$. Cosmic ray protons are expected to loose their energy while propagating from redshift $z_i$ to $z$. 
The redshift evolution of velocity $v=\beta c$ of cosmic ray particles is governed by three processes, (i) the collision with free $e^-$, (ii) ionization of neutral hydrogen, and (iii) the adiabatic expansion of the universe, and is given by \citep{schlickeiser2002cosmic},
\begin{eqnarray}
    \frac{d\beta}{dz}& = & \frac{1}{H(z)\,(1+z)} \biggl[ 3.27 \times 10^{-16} n_e(z) \frac{(1-\beta^2)^{3/2}}{\beta} \frac{\beta^2}{x_m^3 + \beta^3} \nonumber \\ && ~~~~ + 1.94 \times 10^{-16} n_{\rm HI}(z) \frac{(1-\beta^2)^{3/2}}{\beta} \frac{2\beta^2}{(0.01)^{3} + 2\beta^3} \nonumber \\ &&~~~~ \times \{1 + 0.0185 \log \beta \, \Theta(\beta - 0.01)\} \biggr] + \frac{\beta (1-\beta^2)}{(1+z)}
    \label{eq:db_dz}
\end{eqnarray}
where, $n_e(z)$ and $n_{\rm HI}(z)$ are the electron and neutral hydrogen densities respectively in units of $\rm cm^{-3}$, $x_m = 0.0286 (T_g/(2\times10^6\, \rm K))^{1/2}$ and $\Theta$ is the Heaviside step function. 

We note that in Eq.~\ref{eq:db_dz} the two terms inside the square brackets are due to the collision with free electron and ionisation of neutral hydrogen atom respectively, both of which deposit energy to the IGM and contribute to heating. 
In case of a collision with free electron the complete energy loss by a cosmic ray proton becomes the thermal energy of the IGM. 
 
However, when the cosmic ray protons interact with the neutral inter-galactic medium, it may results in primary ionization or excitation to a discrete level and the entire energy of cosmic ray proton does not get transferred to the free electron.  It is shown that the number of primary ion formation is almost proportional to the average loss by collision. In fact, one such ion pair gets formed if the primary loss is about $32$~eV \citep[discussed in sec. 5.3.11 in][]{schlickeiser2002cosmic}. The total energy loss rate is then obtained by multiplying the number density of cosmic ray ($N_{\rm CR}$) with average energy loss and divided by $32$ eV. It is shown that for each ionization process, $\Delta Q \simeq 20$~eV gets deposited as heat \citep{spitzer_1969, Goldsmith_1978}. This leads to a factor $5/8$ in the calculation of energy deposition by cosmic rays.
Hence, the heating rate of the IGM by cosmic ray particles due to ionization can be written as,
\begin{equation}
    \Gamma_{\rm CR}(z, z_0) = \frac{5}{8} \eta_1 \eta_2 \eta_3 \int \frac{dE(\beta)}{dz} N_{\rm CR}(\beta, z, z_0) \, d\beta ,
    \label{eq:gamma_CR}
\end{equation}
where, $\eta_1 = 5/3$ occurs due to the secondary ionization by $e^-$ which produces during ionization by primary cosmic ray particles, $\eta_2 \simeq 1.17$ accounts the $10\%$ He abundance in the IGM, and $\eta_3 \simeq 1.43$ takes into account the contribution of heavy cosmic ray nuclei and cosmic ray electrons ($e^-$), positrons ($e^+$) \citep{schlickeiser2002cosmic}.
These energy depositions due to the cosmic ray particles are incorporated in the temperature evolution of IGM i.e. in Eq.~\ref{eq:Tg} and as we will see they can contribute as a major heating source of IGM during the cosmic dawn.

As already mentioned, the low energy protons that take part in collision and ionization do not escape from the massive atomic cooling Pop~II galaxies.
Only the high energy protons can escape and may interact with high redshift IGM. It should be noted that major part of the cosmic ray energy is carried by the high energy protons ($\sim 1$~GeV).
If a sufficient magnetic field ($B$) is present in the IGM, these high energy cosmic ray particles can gyrate along the magnetic field lines and generate Alfv\'en waves. When these waves get damped, the energy is transferred to the thermal gas \citep{Kulsrud_1969, Skilling1975, Bell1978, Kulsrud_2004}. The energy deposition rate via this Alfv\'en wave generation is $|v_A.\nabla P_c|$, where $v_A= B/\sqrt{4\pi \rho}$ is the Alfv\'en velocity ($\rho $ is the plasma density) and $\nabla P_c$ is the cosmic ray pressure gradient.
Thus the time scale ($t_{\rm CR}$) for this process to influence the IGM of temperature $T_g$ can be calculated by
\begin{equation}
    t_{\rm CR}=\dfrac{3n k_B T_g}{2|v_A.\nabla P_c|}.
\end{equation}
This time scale should be compared to the Hubble time ($t_H$) if it can heat the IGM. Putting some reasonable numbers we find \citep[also see][]{Samui_2018},
\begin{multline}
    \frac{t_{\rm CR}}{t_H} \approx 0.16 \left(\frac{h}{0.7}\right)^4 \left(\frac{\Omega_{\rm m}}{0.3}\right)^{1/2} \left(\frac{1+z}{16} \right)^4 \left(\frac{T_g}{10 \,\rm K} \right) \left(\frac{0.1\,{\rm nG}}{B_0} \right) \\ 
    \times \left( \frac{5 \times 10^{-5}\, {\rm eV/cm^3}}{E_{\rm CR}}\right)
     \left( \frac{L}{0.01\, {\rm Mpc}}\right) ,
\end{multline}
where, $L$ is the physical distance between galaxies and we have taken $L = 0.01$ Mpc as the average separation of typical galaxies at $z=15$.

{Thus we can see if a primordial magnetic field of present day value, $B_0 = 0.1$~nG \citep{Minoda19} is present,} and cosmic rays have energy density of $E_{\rm CR} = 5 \times 10^{-5}\, {\rm eV/cm^3}$ (this can easily be seen from Eq.~\ref{eq:E_CR} using SFRD at $z=15$), the cosmic rays can dissipate their energy to the IGM within Hubble time to influence the IGM temperature of $10$~K at $z\sim15$. Hence, this magnetosonic transfer of energy due to cosmic rays is an important source of IGM heating. To see its influence we assume a fraction {$Q_{\rm CR,II}$} of total cosmic ray energy density is transferred as the thermal energy of the IGM via the Alfv\'en waves from the high energy protons that escape from Pop~II galaxies. 
{It has been shown in \citet{Samui_2018} that if only 10-20\% of cosmic rays energy can be transferred to the IGM it can significantly alter the thermal history of the IGM in the redshift range 2-4. Thus it is natural to consider the same for the thermal history of the IGM during the cosmic dawn. }

 {Here, we would like to mention that cosmic ray electrons from first sources can, in principle, generate radio background through their interactions with intergalactic  magnetic field and alter the global 21-cm signal \citep{Jana_2019}. However, we expect the resulting synchrotron radiation is expected to be insignificant during cosmic dawn given the current upper limit on the intergalactic magnetic field ($B_0 \lesssim 0.1$ nG) \citep{Minoda19, Bera_2020}.}


\subsection{Dark matter - baryon interaction}
\label{sec:DM_th}

As already mentioned, the unusual strong absorption found by EDGES collaboration can possibly be explained by the interaction between dark matter particles and baryons \citep{Barkana18Nature}, and we also consider that here.
{We adopt the dark matter-baryon interaction model from \citet{Munoz15} where the interaction is Rutherford like and the interaction cross-section varies with the dark matter-baryon relative velocity as $\sigma = \sigma_0 v^{-4}$. The dark matter-baryon interaction models are highly constrained by structure formation \citep{Boehm2005}, primordial nucleosynthesis and cosmic rays \citep{Cyburt2002}, CMB anisotropy \citep{Cora2014}, spectral distortions \citet{Yacine2015, James2017}, galaxy clusters \citet{Chuzhoy2006, Hu2008, Qin2001}, gravitational lensing \citet{Natarajan2002, Markevitch2004}, thermal history of the intergalactic medium \citet{Julian2017, Cirelli2009}, 21-cm observations \citep{Tashiro2014}, and so on.
We note that our model satisfy the current limits \citep{Xu2018, Slatyer2018} given on the elastic scattering cross-section between dark matter and baryons using the measurements of CMB temperature and polarization power spectra by the Planck satellite \citep{planck15}, and the Lyman-$\alpha$ forest flux power spectrum by the Sloan Digital Sky Survey (SDSS) \citep{SDSS2004}. The energy transfer rate from dark matters to baryons due to such interactions can be modelled as \citep[see][for the equations that we used to model the dark matter-baryon interaction]{Munoz15, Datta20},}
\begin{multline}
    \frac{dQ_b}{dt} = \frac{2 m_b \rho_{\chi} \sigma_0 e^{-r^{2} / 2} (T_{\chi} - T_g) k_B c^4}{(m_b + m_{\chi})^2 \sqrt{2\pi} u^{3}_{\rm th}} \\
    + \frac{\rho_{\chi}}{\rho_m} \frac{m_{\chi} m_b}{m_{\chi} + m_b} V_{\chi b} \frac{D(V_{\chi b})}{c^2} ,
    \label{eq:dQ_b}
\end{multline}
where, $m_\chi$, $m_b$ and $\rho_\chi$, $\rho_b$ are the masses and energy densities of dark matter and baryon respectively. The dark matter temperature is represented as, $T_{\chi}$. The variance of the thermal relative velocity of dark matter and baryon fluids is given by $u_{\rm th}^2 = k_B(T_b/m_b + T_\chi/m_\chi)$, $V_{\chi b}$ is the relative velocity, and $D(V_{\chi b})$ is the drag term \citep[see][for details]{Munoz15, Bera_2020}.
It can be seen from equation~(\ref{eq:dQ_b}) that the first term on the R.H.S is proportional to the temperature difference between two fluids i.e. $(T_{\chi} - T_g)$. This process is expected to cool the IGM as the temperature of dark matter particles are lower than the baryons in the cold dark matter scenario, and the cooling is much faster than the standard adiabatic cooling mechanism. 
The cooling rate is higher for light dark matter particles and higher interaction cross-section $\sigma_0$. The second term on the R.H.S in equation~(\ref{eq:dQ_b}) arises due to the friction between dark matter and baryon fluids as they have different velocities. So, both the fluids get heated up irrespective of their own temperature. {The frictional term is subdominant compared to the first term during cosmic dawn if one consider a suitable choice of mass of the dark matter particles and the interaction cross-section. However, we consider both the terms in our analysis irrespective of their contribution.}
This heating/cooling rate of baryons ($\dot Q_b$) gets coupled to the gas temperature evaluation ($dT_g/dz$) in Eq.~\ref{eq:Tg}.


\subsection{Lyman-{\ensuremath{\alpha}} coupling}
\label{sec:ly_th}

{As soon as the first generation of stars form, they emit photons at a range of frequencies, including at frequencies higher than Ly-$\alpha$ photons. Photons with  frequency between Ly-$\alpha$ and Ly-$\beta$ get redshifted directly into the Ly-$\alpha$ resonance. The photons having frequency between Ly-$\gamma$ and Ly-limit get redshifted into nearest Lyman series resonance and then  excite  ground state hydrogen atoms. 
These excited HI atoms decay back to $1s$ through radiative cascading process and eventually terminate either in  Ly-$\alpha$ photons or in two $2s \xrightarrow{} 1s$ photons.
Subsequently, the singlet and triplet HI hyperfine level population is altered through Ly-$\alpha$ absorption and emission which is known as the Wouthuysen-Field effect. As the HI spin temperature is determined by the relative population of these hyperfine states, $T_s$ gets affected due to the presence of Ly-$\alpha$ photons. This Ly-$\alpha$ mechanism determines the coupling and decoupling of spin temperature $T_s$ to the gas temperature $T_{g}$ which, in turn, determines the onset of 21-cm absorption signal. Thus, we require the Ly-$\alpha$ flux to calculate the 21-cm signal, and we mostly follow \citet{Barkana_2005, Hirata2006} and \citet{Pritchard_2006} in our work. Moreover, we incorporate the normalisation factor, $f_{\rm esc, \alpha}$ which takes care of the uncertainty in the IMF of the first stars and escape of Ly-$\alpha$ photons. Throughout our work, we consider $f_{\rm esc, \alpha}^{\rm III} = 0.1$ and $f_{\rm esc, \alpha}^{\rm II} = 1.0$ in the calculation of Ly-$\alpha$ flux for Pop~III and Pop~II stars respectively.}
The Ly-$\alpha$ photon intensity or the spherically averaged number of photons striking a gas element per unit area, per unit frequency, per unit time, and per steradian is estimated as,
\begin{equation}
    J_{\alpha} = \frac{(1+z)^2}{4 \pi} \sum_{n=2}^{n_{\rm max}} f_{\rm recycle}(n) \int_{z}^{z_{\rm max}(n)} \frac{c dz^{'}}{H(z^{'})} \epsilon(\nu_n^{'},z^{'}) ,
    \label{eq:J_alpha}
\end{equation}
where, the sum is over all available energy states of hydrogen atom.
We consider the excited atoms upto $n=23$ and hence, the sum is truncated at $n_{\rm max} \simeq 23$ to exclude the levels for which the horizon resides within HII region of an isolated galaxy as pointed out in \citet{Pritchard_2006}. As the 21-cm absorption is governed by only the neutral hydrogen, we can safely use the above approximation. Further, $f_{\rm recycle}(n)$ is the probability that a Ly-$n$ photon will be able to generate a Ly-$\alpha$ photon and the values for different levels are tabulated in Table 1 of \citet{Pritchard_2006}. 
Moreover, a photon with frequency $\nu^{'}_n$ emitted at redshift $z'$  gets redshifted to a Ly series photon corresponding to a level $n$ and absorbed by a ground state HI at redshift $z$. Therefore, we can write \citep{Barkana_2005},
\begin{equation}
    \nu_n^{'} = \nu_{\rm LL} (1-n^{-2}) \frac{1+z^{'}}{1+z} ,
\end{equation}
where, $\nu_{\rm LL}$ is the Lyman limit frequency.
If the photon needs to be seen at Ly-$\alpha$ resonance at a redshift $z$, it should have been emitted at a redshift lower than $z_{\rm max}$ where, 
\begin{equation}
    z_{\rm max}(n) = (1 + z) \frac{[1 - (1+n)^{-2}]}{(1-n^{-2})} - 1 .
\end{equation}
In this way, the fluxes of photons which are emitted between consecutive atomic levels get summed up and contribute to the Ly-$\alpha$ flux.
Further, $\epsilon(\nu, z)$ in Eq.~\ref{eq:J_alpha}, is the comoving photon emissivity defined as the number of photons emitted per unit comoving volume, per proper time and frequency, at rest frame frequency $\nu$ and redshift $z$, and can be obtained from the star formation rate density as, 
\begin{equation}
    \epsilon(\nu, z) = \frac{{\rm SFRD}(z)}{m_p} \times \epsilon_b(\nu) ,
\end{equation}
where, $m_p$ is the mass of proton.
The spectral energy distribution function of the sources, $\epsilon_b(\nu)$, is modelled as a power law $\epsilon_b(\nu) \propto \nu^{\alpha_s - 1}$, where the spectral index $\alpha_s = 1.29$ and $0.14$ for Pop~III and Pop~II stars respectively. It is normalised to emit $4800$ photons per baryons between Ly-$\alpha$ and Lyman limit, out of which $2670$ photons are between Ly-$\alpha$ and Ly-$\beta$ for Pop~III stars. The corresponding numbers for Pop~II stars are $9690$ and $6520$ \citep{Barkana_2005, Pritchard_2006}. 

As already mentioned, the spin temperature ($T_s$) is related to the gas kinetic temperature ($T_g$) and CMBR temperature ($T_{\gamma}$) through collisional coupling and Ly-$\alpha$ coupling (Wouthuysen-Field coupling), and can be obtained using \citep{Field_1958},
{\begin{equation}
    T^{-1}_s = \frac{T^{-1}_{\gamma} + x_{\alpha} T^{-1}_{\alpha} + x_c T^{-1}_g}{1 + x_{\alpha} + x_c} .
    \label{eq:Ts}
\end{equation}
Here, $T_{\alpha}$ is the Ly-$\alpha$ color temperature, and due to the very high optical depth of the Ly-$\alpha$ during cosmic dawn, $T_{\alpha}$ approaches $T_g$ very rapidly. So we consider $T_{\alpha} = T_g$ throughout our work and
hence, Eq.~\ref{eq:Ts} can be written as,}
\begin{equation}
    1 - \frac{T_{\gamma}}{T_s} = \frac{x_{\rm tot}}{1+x_{\rm tot}} \left( 1 - \frac{T_{\gamma}}{T_g} \right) ,
\end{equation}
where, $x_{\rm tot} = x_{\alpha} + x_c$, the sum of the coupling coefficients due to scattering of Ly-$\alpha$ photons and atomic collisions respectively. The Wouthuysen-Field coupling coefficient is given by \citep{Wouthuysen_1952, Field_1958},
\begin{equation}
    x_{\alpha} = \frac{16 \pi^2 T_{*} e^2 f_{\alpha}}{27 A_{10} T_{\gamma} m_e c} S_{\alpha} J_{\alpha} ,
    \label{x_a}
\end{equation}
and the collisional coupling coefficient can be calculated as,
\begin{equation}
    x_c = \frac{\kappa^{\rm HH}_{10}(T_g) n_{\rm HI} T_*}{A_{10} T_{\gamma}} ,
    \label{x_c}
\end{equation}
where $f_{\alpha} = 0.4162$ is the oscillator strength for the Ly-$\alpha$ transition. The Ly-$\alpha$ photon intensity $J_{\alpha}$ is calculated using Eq.~\ref{eq:J_alpha}. Further, $S_{\alpha}$ in Eq.~\ref{x_a} is a correction factor of order unity which takes care of the redistribution of photon energies due to the repeated scattering off the thermal distribution of atoms, and we assume it to be $S_{\alpha} = 1$ \citep{Chen_2004}. To calculate the specific rate coefficient $\kappa^{\rm HH}_{10}$, we use the fitting formula $\kappa^{\rm HH}_{10}=3.1\times10^{-17}T_g^{0.357}\exp(-32/T_g) \, \rm{m^3s^{-1}}$ as given in \citet{pritchard12}.
Also, $T_* = h_p \nu_e/k_B = 0.068$~K is the characteristic temperature for the HI 21-cm transition. Note that, the Ly-${\alpha}$ coupling is the dominant coupling mechanism during cosmic dawn, whereas the collisional coupling dominates during dark ages.

\section{Results}
\label{main_sec:results}

In this section, we present our results of global 21-cm signal highlighting the heating due to cosmic ray protons generated through supernova explosions from very early Pop~III and Pop~II stars. For better comprehension, we list below the default parameters that we used in our work.
The reason for considering such parameter values has already been discussed in previous sections.

Pop~III model :
\begin{itemize}
    \item {Each minihalo generates a single Pop~III star of $145$~$M_{\odot}$ which explodes as a core-collapse supernovae, so $f_{\rm SN,III} = 1/145$.}
    \item {The explosion energy of a Pop~III supernova is taken as $E_{\rm SN,III} = 10^{52}$~erg.}
    \item {{A fraction of SNe kinetic energy, $\epsilon_{\rm III} = 0.06$ gets utilised to accelerate the cosmic rays. This value is chosen in order to match the EDGES deep absorption profile, though it is a free parameter in our model and we show the variation of this parameter in the later section.}}
    \item {The slope, $q$ of the cosmic ray spectra (Eq.~\ref{eq:p_spectra}) generated by the supernova, is taken as $2.2$.}
\end{itemize}

Pop~II model : 
\begin{itemize}
    \item {We consider one supernova explosion per $50$~$M_{\odot}$ of star formation, i.e. $f_{\rm SN,II} = 0.02$}
    \item {The typical energy of Pop~II supernova is $E_{\rm SN,II} = 10^{51}$~erg.}
    \item {The fraction of SNe kinetic energy carried by the cosmic rays is $\epsilon_{\rm II} = 0.15$.}
    \item {{The amount of cosmic ray energy that gets transferred to the IGM and heats the surrounding gas is $Q_{\rm CR,II} = 0.15$. This value is chosen to explain the sharp rise of the EDGES profile. The variation of this parameter is shown in the Fig.~\ref{fig:T21_noDM}}}.
\end{itemize}

These are the default parameter values adopted from \citet{Mebane_2018} and \citet{Samui_2014} for Pop~III and Pop~II star formation models respectively, and from \citet{schlickeiser2002cosmic} $\&$ \citet{Samui_2005} for cosmic rays energy deposition. These parameters are chosen keeping in mind the EDGES observation. We also show results by varying parameters such as $q, \epsilon_{\rm III}, \epsilon_{\rm II}$ {and $Q_{\rm CR,II}$. As the impact of $\epsilon_{\rm II}$ and $Q_{\rm CR,II}$ are similar, we don't vary them separately rather we vary only $\epsilon_{\rm II}$, and represent it as $\epsilon_{\rm II}$ and/or $Q_{\rm CR,II}$.}

\subsection{Lyman-Werner background and star formation rate densities}

\begin{figure}
\centering
\begin{subfigure}[b] {0.45\textwidth}
   \includegraphics[width=\columnwidth]{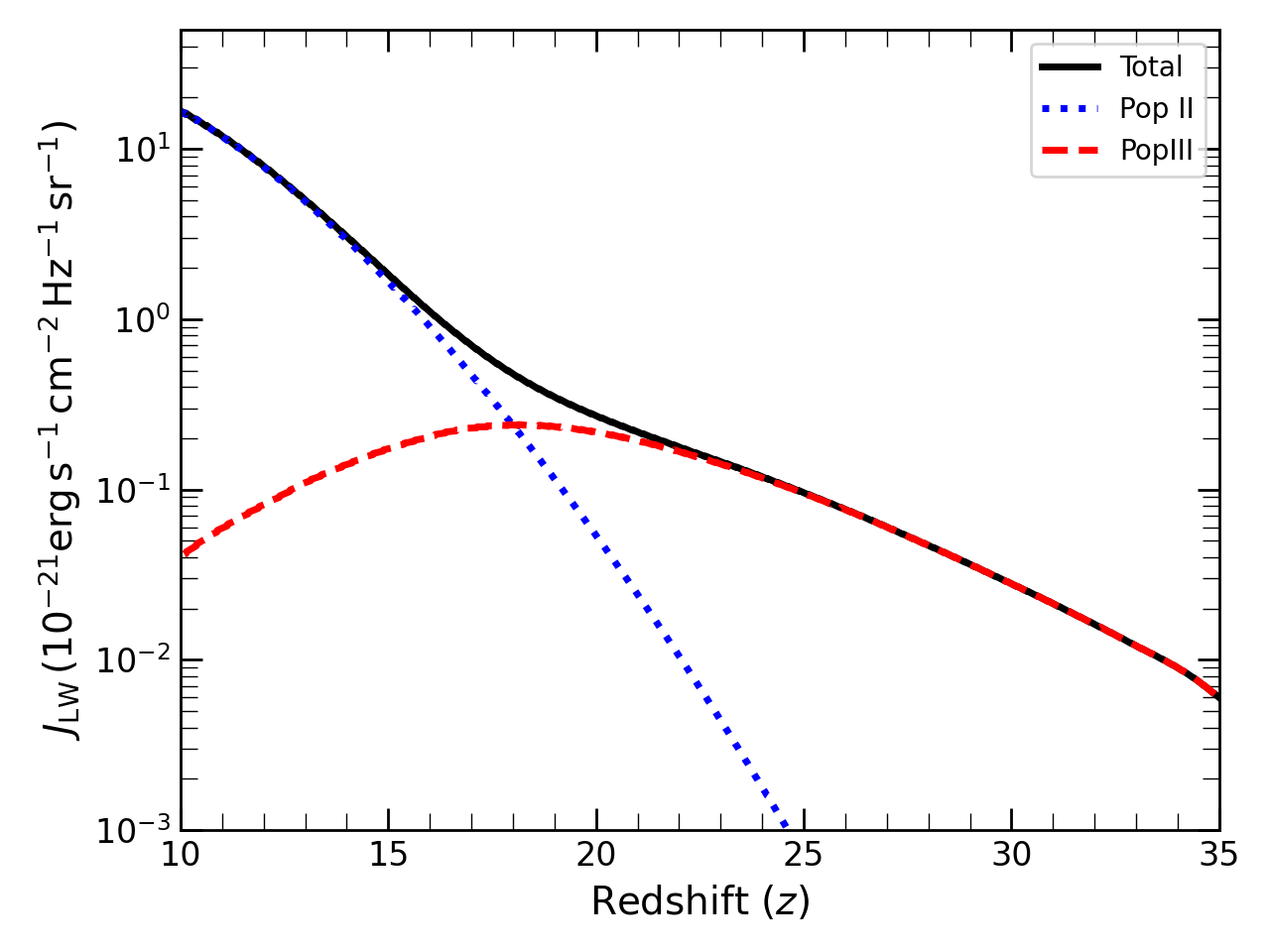}
   \caption{}
   \label{fig:J_LW} 
\end{subfigure}

\begin{subfigure}[b] {0.45\textwidth}
   \includegraphics[width=\columnwidth]{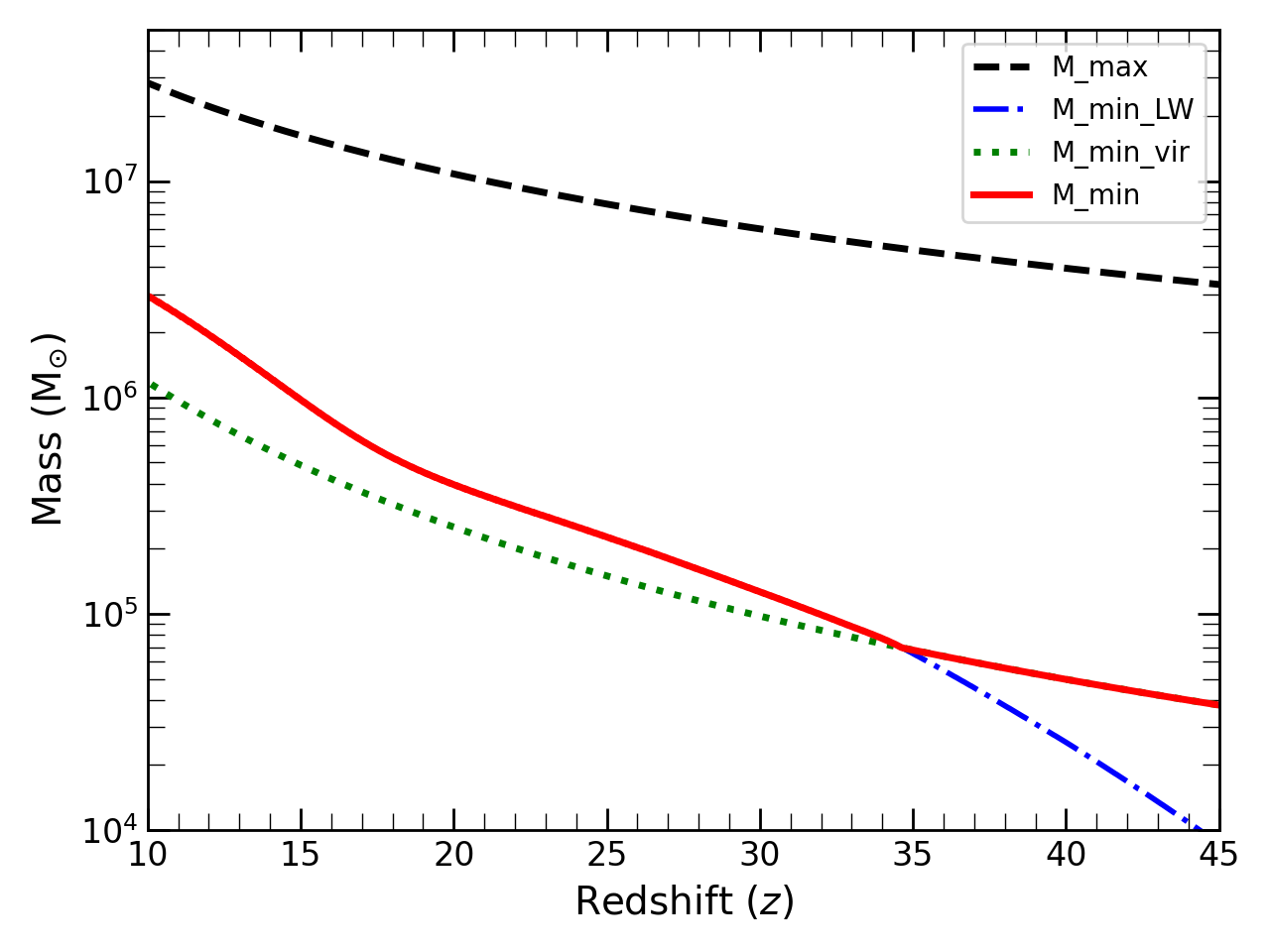}
   \caption{}
   \label{fig:mass_plot}
\end{subfigure}

\caption{(a) The total Lyman-Werner flux, $J_{\rm LW}$, in units of $10^{-21} \, \rm erg \, s^{-1} cm^{-2} Hz^{-1} sr^{-1}$ at different redshift $z$ is plotted by black solid curve. The contribution from Pop~III and Pop~II stars are also plotted separately by red dashed and blue dotted curves respectively. (b) The black dashed curve represents the maximum halo mass corresponding to the molecular cooling cut-off for Pop~III star formation. The green curve represents the minimum halo mass required for Pop~III stars to form, whereas the blue dash dotted one is the same considering Lyman-Werner feedback. At each redshift, the larger among the above two acts as the minimum halo mass at that redshift, represented here by the red solid curve.}
\end{figure}

We begin by showing our results on star formation rate.
As already discussed, we consider both metal free Pop~III stars and metal enriched Pop~II stars. We begin our numerical calculation from redshift $z\sim 50$. Pop~III star formation gets suppressed due to the presence of Lyman-Werner (LW) photons which, in turn, reduces the emission of LW photons from Pop~III stars itself. Therefore, the LW flux and Pop~III star formation rate density are calculated simultaneously and self consistently.
The resulted LW specific intensity, $J_{\rm LW}$ as a function of redshift $z$ is shown in Fig.~\ref{fig:J_LW}. The total LW flux is represented by the black solid curve, whereas the contribution from Pop~III and Pop~II stars are shown separately by red dashed and blue dotted curves respectively. It is clear from the Fig.~\ref{fig:J_LW} that the LW contribution is dominated by Pop~III stars in early epochs. However, the contribution is taken over by the LW photons coming from the Pop~II stars at redshift $z \sim 18$. This rapid increase in LW background dissociates more and more H$_2$ molecules and, as a result, the minimum mass required to form Pop~III stars increases as can be seen in Fig.~\ref{fig:mass_plot}. Here the minimum (red solid) and maximum masses (black dashed curve) of dark matter halos that could form Pop~III stars are plotted at different redshifts. The maximum halo mass for Pop~III star formation corresponds to the atomic cooling cut-off temperature, i.e. $10^4$~K. The minimum mass required for Pop~III star formation depends upon two factors, i) halos which satisfies the criteria $f_{\rm H_2} > f_{\rm crit,H_2}$ (discussed in Section~\ref{subsec:PopIII}) at each redshift (green dotted curve in Fig.~\ref{fig:mass_plot}), ii) the minimum mass corresponding to Eq.~\ref{eq:LW_mass} determined by the LW flux (Eq.~\ref{eq:J_LW}) and plotted as blue dash-dotted curve in Fig~\ref{fig:mass_plot}. At any redshift, the larger halo mass among the above two is considered to be the minimum halo mass (shown by red solid curve) for the calculation of Pop~III SFR density. As can be seen from the figure the minimum halo mass can be determined from the Eq.~\ref{eq:f_cH2} at redshifts i.e. $z \gtrsim 35$. However, in later epochs the minimum mass is set by the LW background flux. 

\begin{figure}
	\includegraphics[width=\columnwidth]{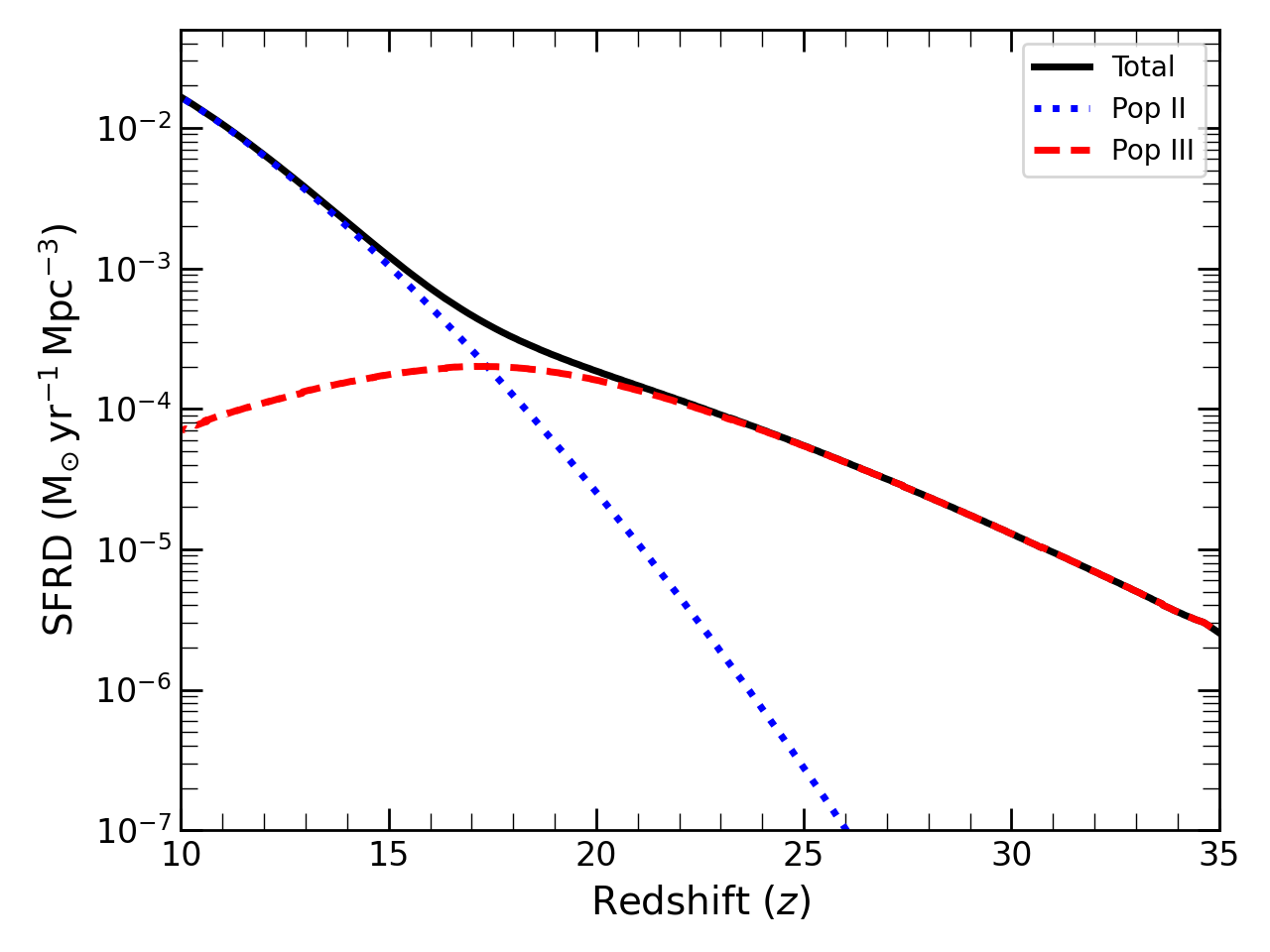}
    \caption{The total star formation rate density is represented by the black solid curve whereas the contribution of Pop~III and Pop~II stars are denoted by the red dashed and blue dotted curves respectively. 
    }
    \label{fig:SFR}
\end{figure}
Note that, around $z\sim 17$ the total LW background increases very rapidly. This suppresses the Pop~III star formation rate and, as a consequence, the production of LW photons from Pop~III stars decreases at redshifts $z \lesssim 17$. It can be seen in Fig.~\ref{fig:SFR} where the star formation rate density as obtained in our model has been plotted.
Fig.~\ref{fig:SFR} shows the Pop~III (red dashed curve), Pop~II (blue dotted curve), and total (black solid curve) star formation rate densities.
We see in Fig.~\ref{fig:SFR} that the Pop~III star formation rate which initially rises with decreasing redshift dominates over the Pop~II star formation rate at redshifts $z \gtrsim 17$. As the LW background due to Pop~II stars increases, it narrows down the halo mass range that can form Pop~III stars. As a consequence, the Pop~III star formation decreases at redshift below $z \sim 17$ and the Pop~II star formation rate takes over the Pop~III at redshifts $z \lesssim 17$. 
Note that, the exact epoch of domination of Pop~II over Pop~III stars, in principle, depends, on several parameters such as, the mass of the stars formed in Pop~III halos, the form of halo mass function one assumes, star formation efficiencies of Pop~III and Pop~II stars etc. However, we do not expect the qualitative behaviour of the above results to change significantly.

\subsection{Ly-{\ensuremath{\alpha}} coupling and resulting spin temperature}

\begin{figure}
	\includegraphics[width=\columnwidth]{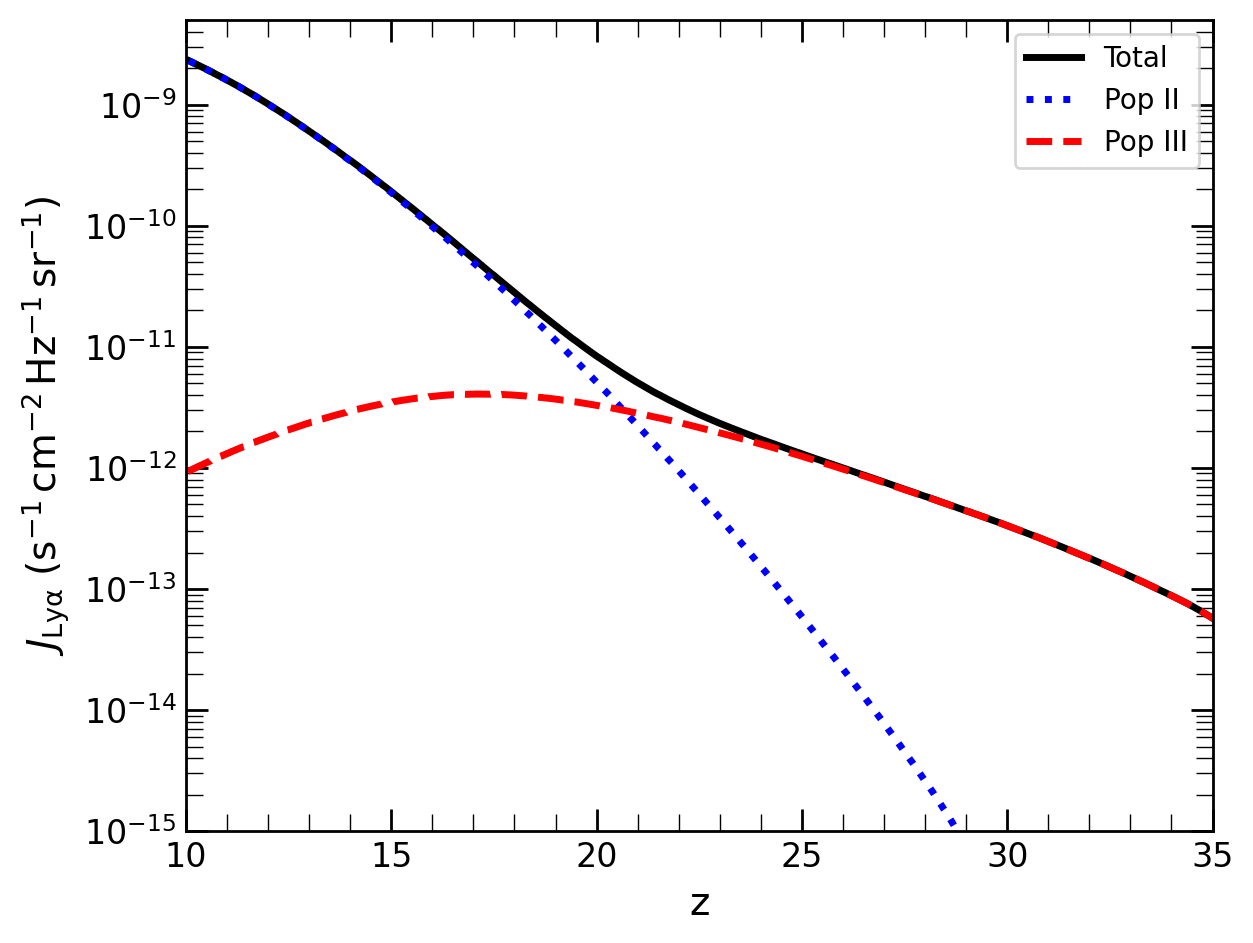}
    \caption{Lyman-$\alpha$ flux as a function of redshift $z$ that arises from both Pop~III and Pop~II stars. Black solid curve denotes the total Ly-$\alpha$ flux and the representation of separate contributions of Pop~III and Pop~II stars is same as Fig.~\ref{fig:SFR}.}
    \label{fig:J_alpha}
\end{figure}

In this section, we discuss the Lyman-$\alpha$ flux resulting from the Pop~III and Pop~II star formation and the corresponding coupling of hydrogen spin temperature and IGM temperature. Fig.~\ref{fig:J_alpha} shows the Ly-$\alpha$ specific intensity as a function of redshift $z$. The total Ly-$\alpha$ photon flux is shown by the black solid curve, whereas the contributions from the Pop~III and Pop~II stars are also plotted by the red dashed and blue dotted curves respectively. The Ly-$\alpha$ flux is plotted in units of $\rm s^{-1} cm^{-2} Hz^{-1} sr^{-1}$. Our model accounts for the recycling of Ly-$n$ photons that ultimately cascade down to Ly-${\alpha}$ photons. We see that the contribution from the Pop~III stars dominates the total Ly-$\alpha$ intensity at redshift $z \gtrsim 20$. Due to strong Lyman-Werner feedback at lower redshifts, star formation enabled through molecular-cooling gets suppressed and, as a consequence, Ly-$\alpha$ emission from these population decreases gradually at lower redshifts. 

\begin{figure*}
	\includegraphics[width=\columnwidth]{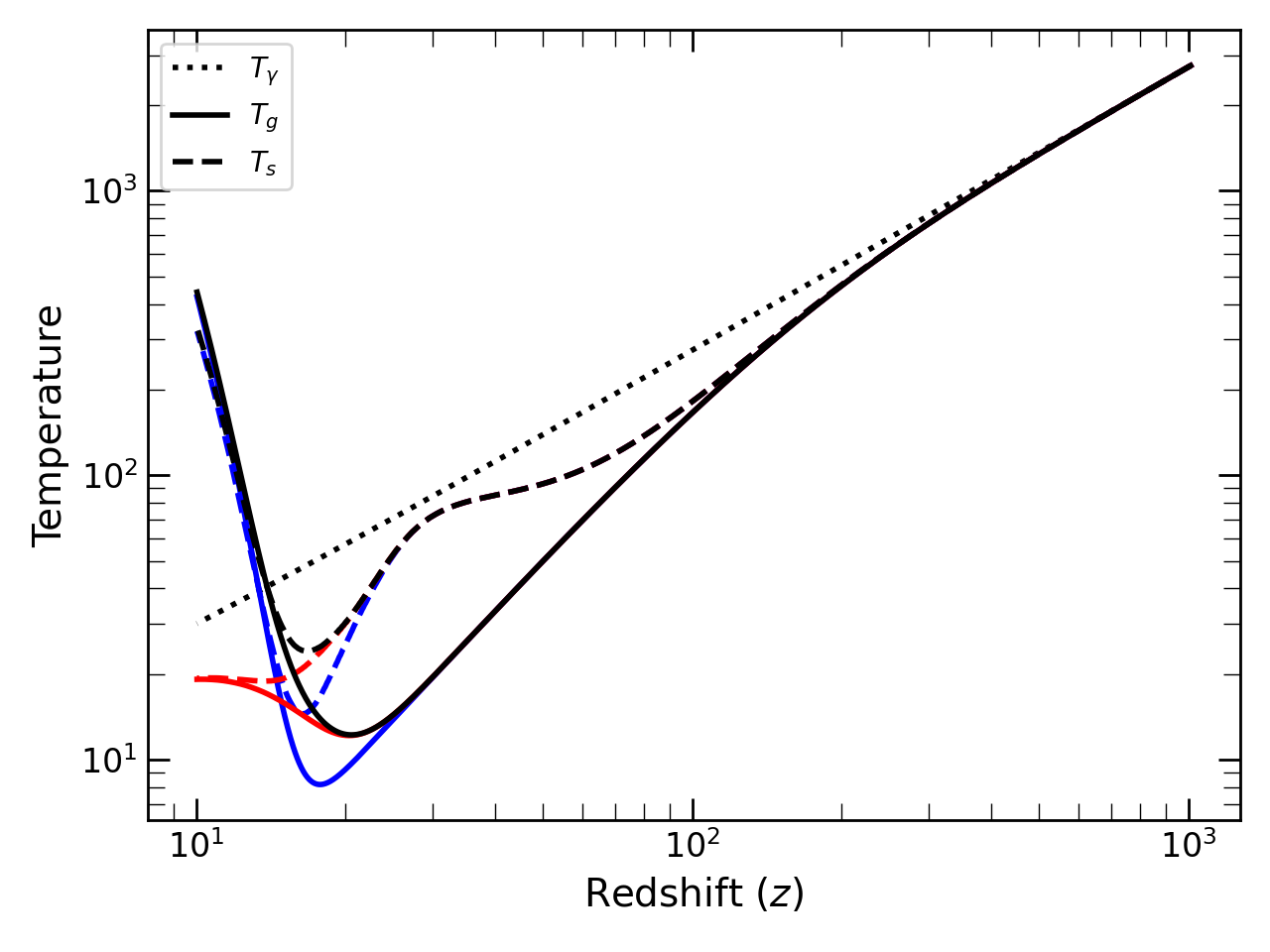}
	\includegraphics[width=\columnwidth]{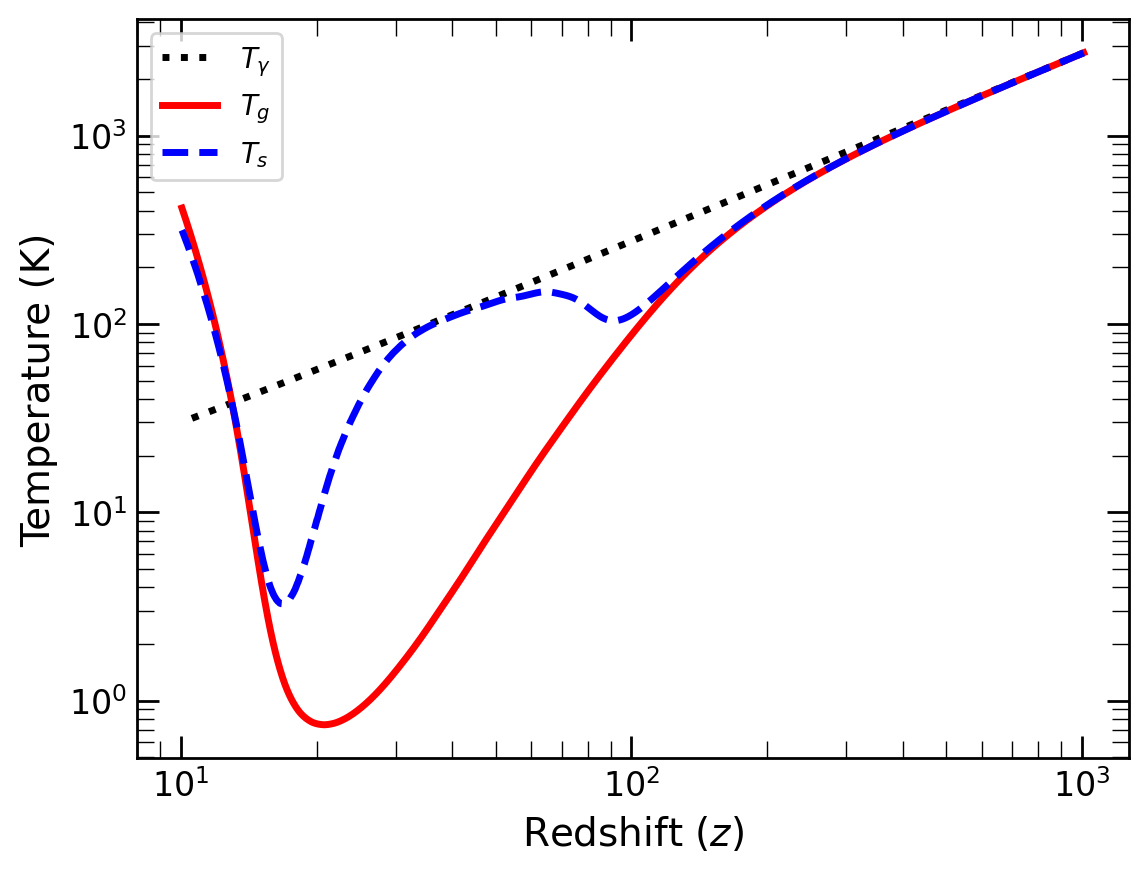}
    \caption{{{\bf Left panel:} The evolution of the gas kinetic temperature, $T_g$ with redshift is shown by black solid curve along with the spin temperature, $T_s$, and CMBR temperature $T_{\gamma}$ by black dashed and black dotted curves respectively. The blue solid and dashed curves represent the evolution in $T_g$ and $T_s$ respectively for considering the contribution of cosmic rays from Pop~II stars only and the red curves represent the same for considering Pop~III stars only.} {\bf Right panel:} The evolution of the gas kinetic temperature, $T_g$ with redshift is shown by red solid curve along with the spin temperature, $T_s$, and CMBR temperature $T_{\gamma}$ by blue dashed and black dotted curves respectively. The dark matter-baryon interaction is included here for excess cooling mechanism keeping in mind the EDGES deep absorption signal. The dark matter mass, and interaction cross-section $(m_{\chi}/{\rm Gev}, \sigma_{45}) = (0.1, 2)$ are considered for this particular plot.}
    \label{fig:Ts_Tg}
\end{figure*}

As soon as the first sources appear, Ly-$\alpha$ photons from these sources help the spin temperature $T_s$ to decouple from the CMBR temperature and starts coupling with the gas temperature. This can be seen from Fig.~\ref{fig:Ts_Tg}, where we have plotted the CMBR temperature, spin temperature and gas temperature starting from the recombination redshift $z = 1010$ {without DM-b interaction (left panel) and with DM-b interaction (right panel)}. It is clear from the figure that the decoupling begins at a redshift as early as $z \sim 30$ in our model. The spin temperature, $T_s$ then gradually approaches to the IGM kinetic temperature, $T_g$, and gets fully coupled at redshift $z \approx 17$. Since the Pop~II stars dominate the Ly-$\alpha$ photon budget at redshift $z \lesssim 20$, the coupling is mainly determined by this population {as shown in the left panel of Fig.~\ref{fig:Ts_Tg}, where we plotted the spin temperatures by blue (Pop~II), red (Pop~III) and black (Pop~II + Pop~III) dashed curves}. We note that, along with the standard adiabatic cooling of the IGM prior to the heating, we also consider dark matter-baryon interaction in our work. {The right panel of }Fig.~\ref{fig:Ts_Tg} shows result for dark matter mass $m_{\chi}=0.1$ GeV and the interaction cross-section $\sigma_{45}=\frac{\sigma_0}{10^{-45} \, {\rm m^2}} =2$. This interaction  helps to cool the IGM  faster compared to the standard adiabatic cooling. This scenario is motivated by the EDGES observation that shows a strong absorption profile of {-0.5$~K$}. Redshifts of the decoupling of $T_s$ from the CMBR temperature and coupling to the IGM kinetic temperature can change to some extent depending on the Ly-$\alpha$ escape fraction into the IGM and redshift evolution of the IGM kinetic temperature. However, as we focus on the impact of cosmic rays on heating of the IGM here, we defer this discussion for future works. 

\begin{figure}
	\includegraphics[width=\columnwidth]{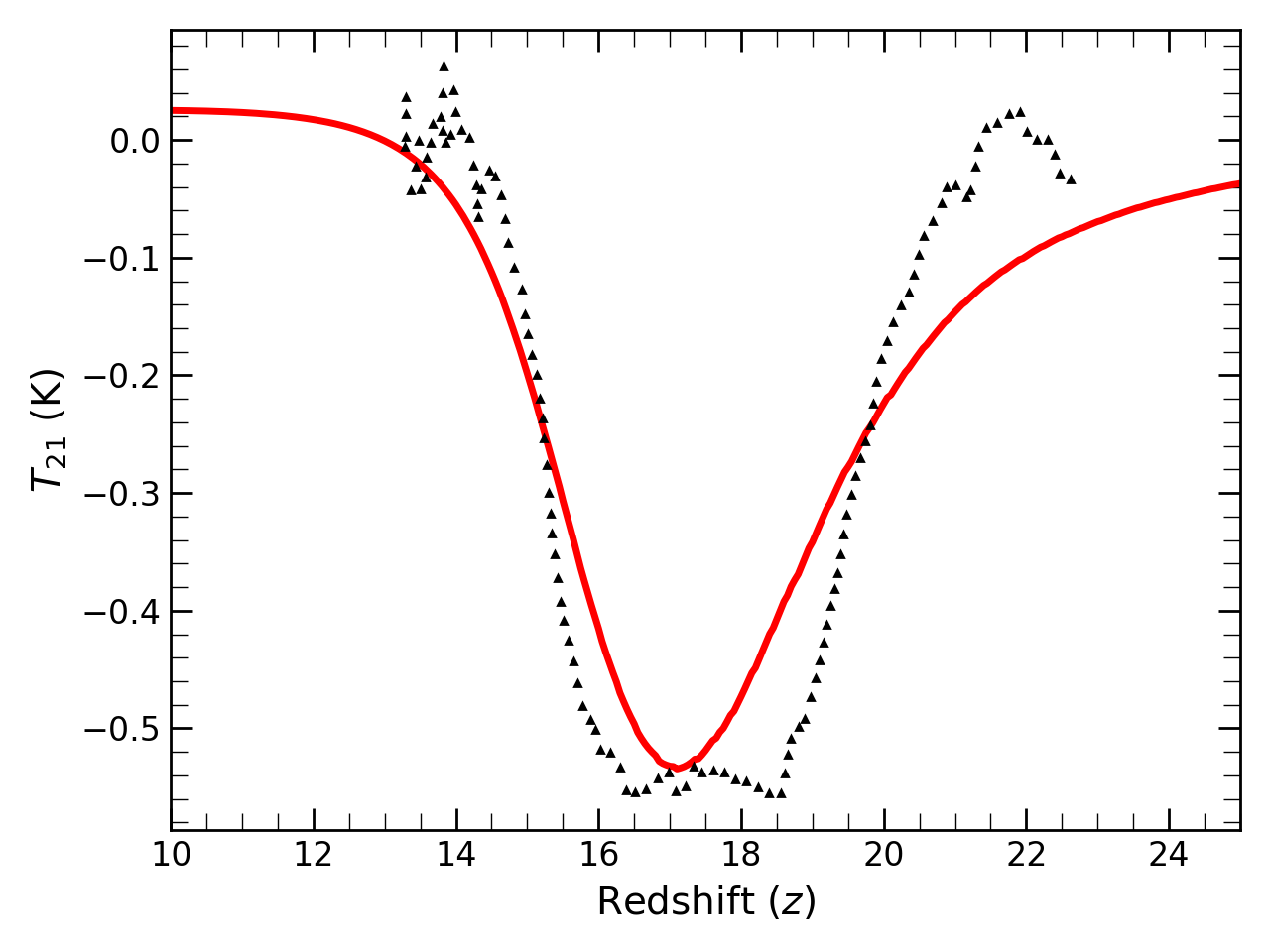}
    \caption{The brightness temperature, $T_{21}$, resulting from the temperatures shown in Fig.~\ref{fig:Ts_Tg} is represented here by red solid curve. This plot is generated considering Ly-$\alpha$ coupling and cosmic ray heating with efficiencies, $\epsilon_{\rm III} = 0.06$, and {$\epsilon_{\rm II}$ and/or $Q_{\rm II} = 0.15$} for cosmic rays produced in Pop~III and Pop~II stars respectively. The dark matter-baryon interaction parameters, $(m_{\chi}/{\rm Gev}, \sigma_{45})$ are same as in Fig. \ref{fig:Ts_Tg}. For reference, the measured $T_{21}$ signal by the EDGES is shown here by black triangles.}
    \label{fig:T21}
\end{figure}

\subsection{Impact of cosmic rays heating on IGM temperature}

One of the major aims of this work is to explore cosmic rays from first generation of stars as a source of IGM heating during the cosmic dawn. This subsection discusses our results on that.  As already discussed, we consider the collisional and ionization interaction as well as the magnetosonic interaction through which the energy gets transferred from cosmic rays to the IGM that leads to the increase in kinetic temperature of the IGM.
In Fig.~\ref{fig:Ts_Tg}, we have plotted the resulting IGM gas temperature by the red solid curve. 
Initially upto redshift $z \gtrsim 200$, the gas kinetic temperature $T_g$ and CMBR temperature $T_{\gamma}$ follow each other. This is enabled through Compton scattering process. Afterwards $T_g$ cools faster than $T_{\gamma}$ due to adiabatic cooling and dark matter-baryon interaction that we consider. Fig.~\ref{fig:Ts_Tg} shows results for the dark matter mass of $m_{\chi} = 0.1$ GeV, and interaction cross-section of $\sigma_{45} = 2 $ that resulted in the sharp fall of IGM temperature. Further, note that, due to the collisional coupling hydrogen spin temperature follows the IGM temperature upto $z \gtrsim 100$. Afterwards the collisional coupling becomes weak due to the lower IGM density and temperature, and hence $T_s$ again starts to follow the CMBR temperature. As discussed before, the spin temperature again decoupled from the CMBR due to the presence of Ly-$\alpha$ photons around $z \sim 30$. Meanwhile the IGM temperature reaches a minimum  at $z \sim 20$. By this time the first generation of stars has produced enough amount of cosmic rays that start heating the surrounding gas and increase the IGM temperature as can be seen from Fig.~\ref{fig:Ts_Tg}.  $T_g$ crosses the CMBR temperature at redshift $z \sim 12$ due to the cosmic ray heating. By this redshift, the Ly-$\alpha$ coupling has completely coupled the spin temperature to gas temperature. The interplay between $T_s$ and $T_{\gamma}$ determines the 21-cm brightness temperature that we discuss in the next section.

\subsection{Global 21-cm signal}

The EDGES observation suggests a mean brightness temperature, $T_{21}$ between $-0.3$~K to $-1.0$~K. As already discussed, this strong absorption can be explained by considering a colder IGM resulting from dark matter-baryon interaction as adopted in our model. Thus, we will first focus our result in the light of EDGES detection. Later, we will also discuss the effect of cosmic ray heating on the global 21-cm signal in the absence of any dark matter-baryon interaction.

In the Fig.~\ref{fig:T21}, we show the estimated global 21-cm signal using $T_g$ and $T_s$ as discussed in the previous sections by solid red line. 
For comparison, we have also plotted the measured profile of $T_{21}$ with black triangles using publicly available data of \citet{EDGES18}.
The absorption feature starting at $z \sim 20$ arises from the coupling of $T_s$ with $T_g$ by Ly-$\alpha$ photons produced by the early generation of Pop~III stars. The deeper absorption of $-0.5$~K is resulting from the dark matter-baryon interaction that cools the IGM temperature upto $\sim 1$~K by $z \sim 17$. Afterwards, the heating due to cosmic ray protons generated from early Pop~III and Pop~II stars increases the IGM temperature sharply. By $z \sim 14$, the IGM temperature exceeds the CMBR temperature due to this cosmic ray heating. This causes the sharp disappearance of the absorption feature in the 21-cm signal as detected by the EDGES collaboration. {Thus we can say that the rising arm of the absorption signal matches reasonably well with the cosmic ray heating along with the dark matter-baryon interaction. }

We note that, for this particular profile, very small amount of energy gets transferred from the cosmic rays to IGM. For instance, we have used $\epsilon_{\rm III} = 0.06$, and {$\epsilon_{\rm II}$ and/or $Q_{\rm II} = 0.15$} for Pop~III and Pop~II stars. 
{In passing we note that, even though we focus our results in light of EDGES detection, there are controversies about the detected signal by EDGES. For example, a recent work by SARAS3 collaboration \citep{Saurabh_2021} has claimed a null detection of the same signal by a different experimental setup. However the cosmic ray heating is a more generic physics that one should consider while modelling the cosmic 21-cm signal. The exotic dark matter-baryon interaction was/is considered only to explain the unusually strong absorption profile detected by EDGES group. Hence, in order to highlight the effect of cosmic ray heating on 21-cm signal, and keeping in mind the controversy regarding the detected signal, we describe the effect of cosmic ray heating on 21-cm signal without considering the exotic dark matter-baryon interaction, in the next section.}

\subsection{21-cm signal without dark matter-baryon interaction}

\begin{figure}
	\includegraphics[width=\columnwidth]{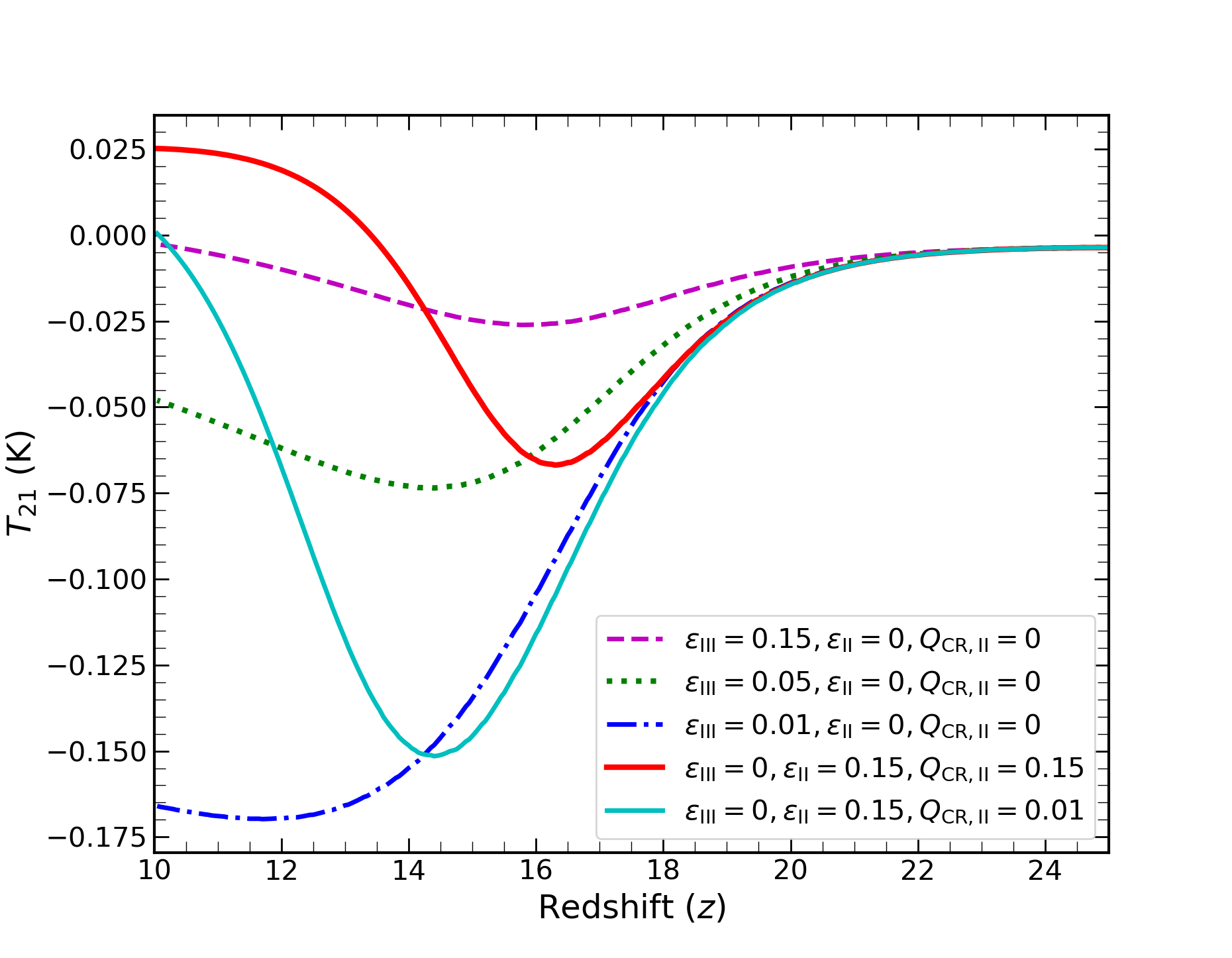}
    \caption{Variations in the global HI 21-cm signal $T_{21}$ due to the various efficiencies of cosmic ray heating are shown here. We kept the spectral index of cosmic ray spectra, $q$  fixed for this plot. All the evolution are done here without considering any exotic mechanism such as dark matter-baryon interaction.}
    \label{fig:T21_noDM}
\end{figure}

In order to understand the contribution of cosmic ray heating from Pop~III and Pop~II stars separately, we vary the efficiency parameters $\epsilon_{\rm III}$, $\epsilon_{\rm II}$ {and the amount of energy transferring from cosmic rays to the IGM, $Q_{\rm CR, II}$} keeping the cosmic ray spectral index $q$ fixed. The resulting brightness temperature, $T_{21}$ as obtained for various $\epsilon_{\rm II}$, $\epsilon_{\rm III}$ {and $Q_{\rm CR,II}$} is shown in Fig.~\ref{fig:T21_noDM}. It is clear from the figure that in all such models, the absorption profile starts at $z \sim 20$ due to the Ly-$\alpha$ coupling. This depends only on the star formation rates of Pop~III and Pop~II stars and resulting Ly-$\alpha$ flux. It is independent of cosmic ray heating as long as the gas temperature remains below the CMBR temperature. However, the depth and width of the absorption spectra highly depend on the heating due to cosmic rays. For example, if the Pop~III SNe are more efficient in accelerating cosmic rays i.e. $\epsilon_{\rm III} = 0.15$, we get a very shallow absorption profile of $\sim -0.025$~{K} as can be seen by magenta dashed curve in Fig.~\ref{fig:T21_noDM}. If the efficiency is even higher, it is likely to wash out any possible 21-cm absorption profile. On the other hand, reducing the efficiency would increase the absorption depth as well as the duration of the absorption as can be seen from the green dotted and blue dash dotted curves where $\epsilon_{\rm III} = 0.05$ \& $0.01$ respectively. Note that, in these three models, we didn't take any contribution of cosmic ray heating by Pop~II stars. Thus we can say that a very small amount of contribution in heating by cosmic rays generated from Pop~III stars ($\epsilon_{\rm III} = 0.01$), would reduce the absorption depth compared to the prediction of the same by any model where no cosmic ray heating is considered. Thus accurately determined global 21-cm signal would constrain the contribution of cosmic ray heating which in turns can put constraint on the nature of Pop~III stars during the cosmic dawn.

Finally, we show the contribution of cosmic ray heating only from Pop~II stars by red solid curve in Fig.~\ref{fig:T21_noDM}, where we assumed $\epsilon_{\rm III} = 0$, $\epsilon_{\rm II} = 0.15$ and $Q_{\rm CR,II} = 0.15$. It is clear from the figure that Pop~II heating is also efficient in reducing the absorption depth as well as the duration of the absorption profile. In this case, we get a maximum absorption of $-0.07$~{K} at $z\sim 16$ which ends by $z\sim 13.5$.
Even for a very small efficiency of $Q_{\rm CR,II} = 0.01$, we get an absorption depth of {$\sim -0.15$~K} as can be seen from cyan solid curve in Fig.~\ref{fig:T21_noDM}. This also increases the duration of the absorption signal. Note that, the impact of $\epsilon_{\rm II}$ is the same as that of $Q_{\rm CR,II}$. Thus even the heating by cosmic rays generated from the Pop~II stars is an important physics that one should consider while modelling the cosmic 21-cm signal. We also note that Pop~III stars have a considerable impact on the global 21-cm signal at redshift  $z\gtrsim 16$ in all models we consider. 
{However, $T_g$, $T_s$, and the resulting absorption profile depends significantly on the assumed parameters of cosmic ray heating that we discuss in the next section.}

\subsection{Variation of model parameters}

\begin{figure*}
    \centerline{\includegraphics[width=\columnwidth]{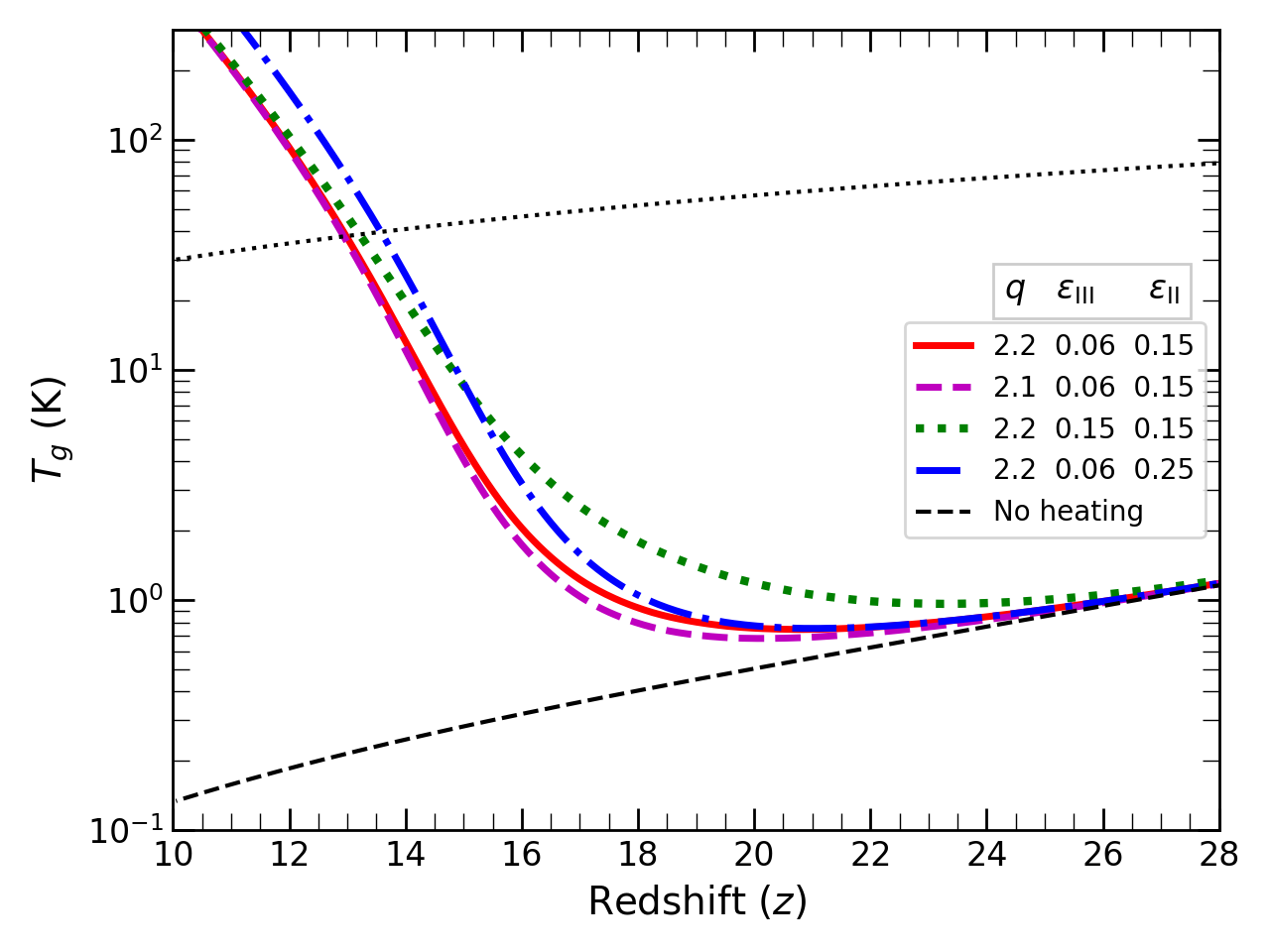}\includegraphics[width=\columnwidth]{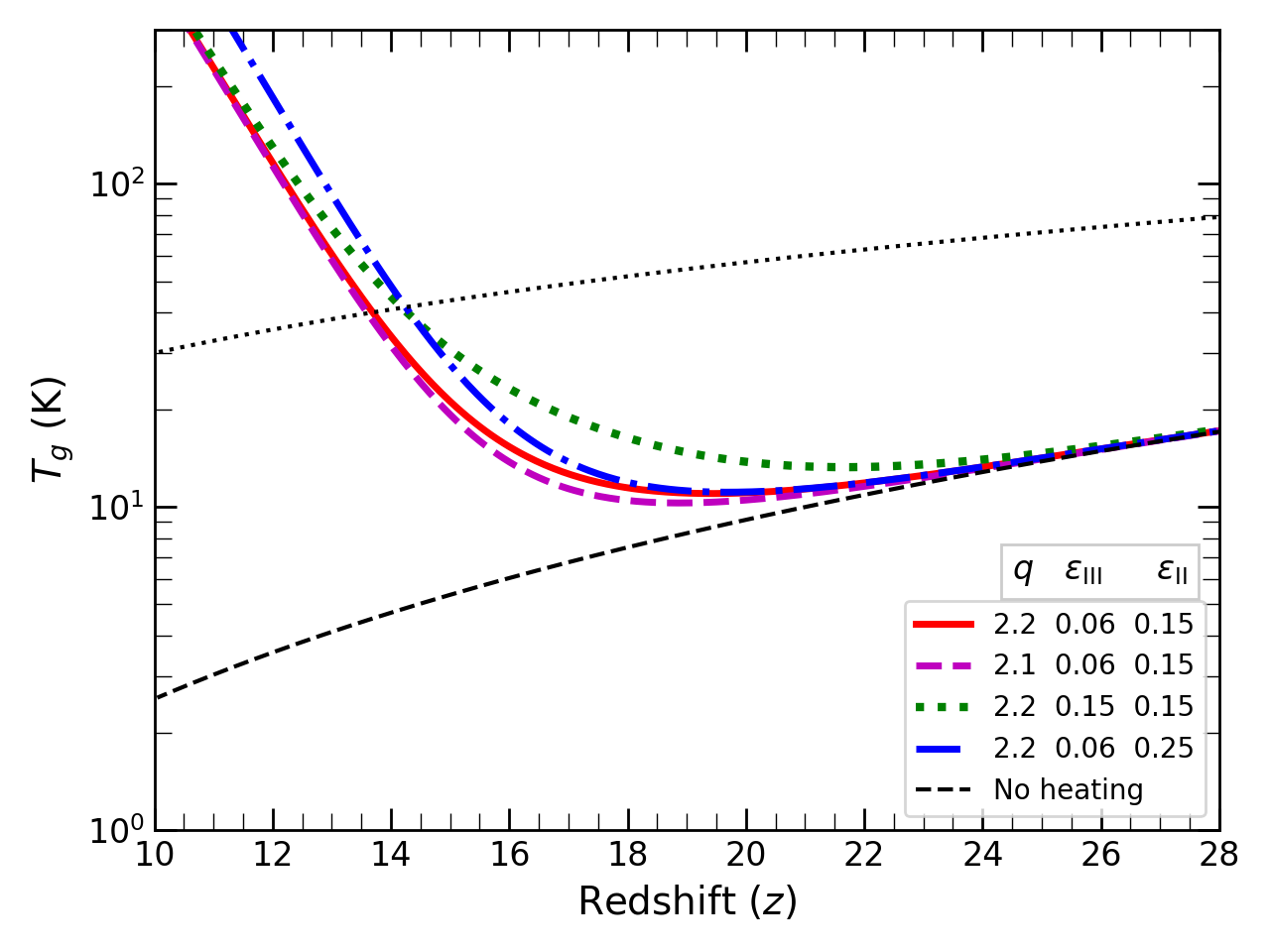}}
    \centerline{\includegraphics[width=\columnwidth]{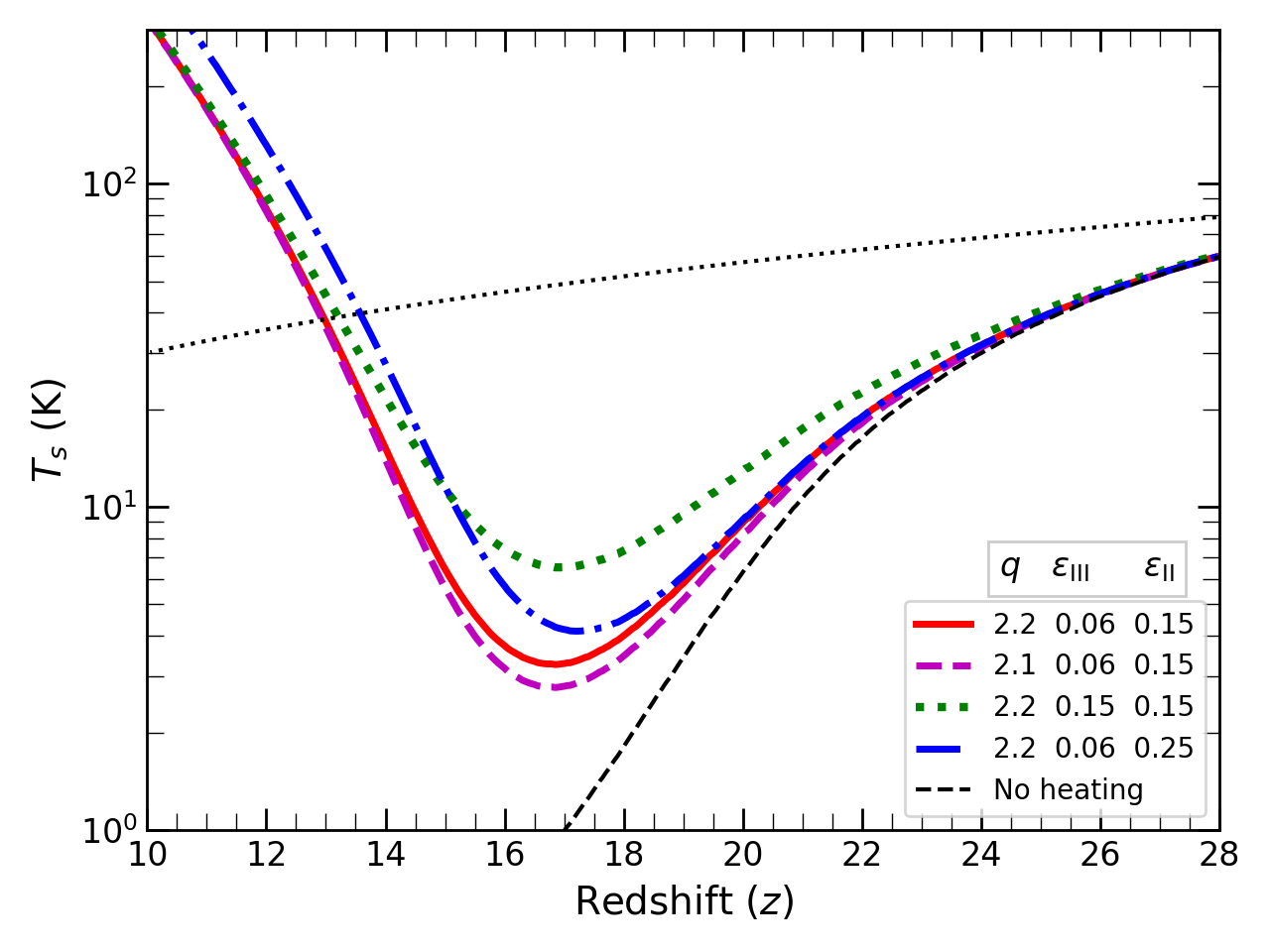}\includegraphics[width=\columnwidth]{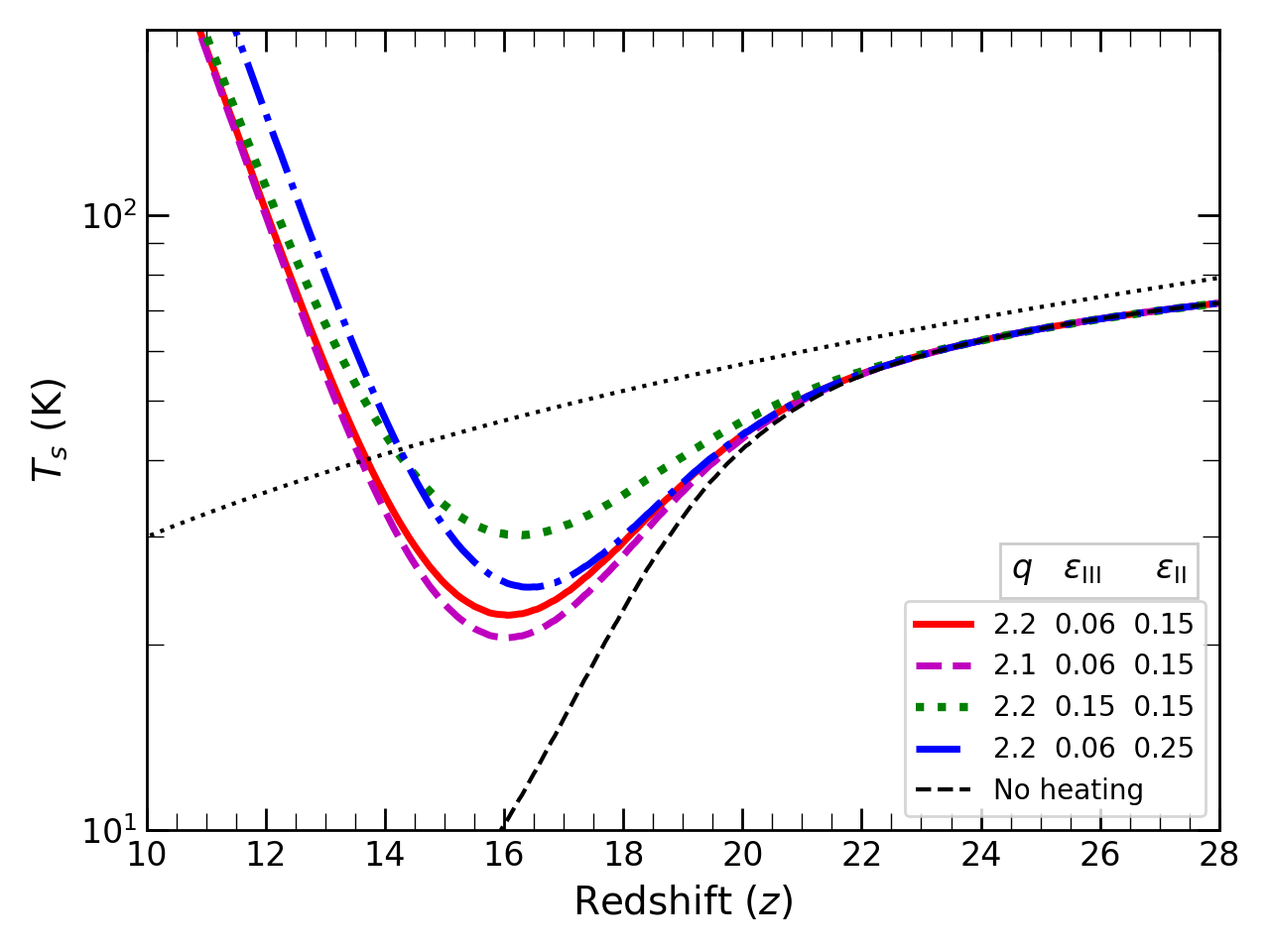}}
    \centerline{\includegraphics[width=\columnwidth]{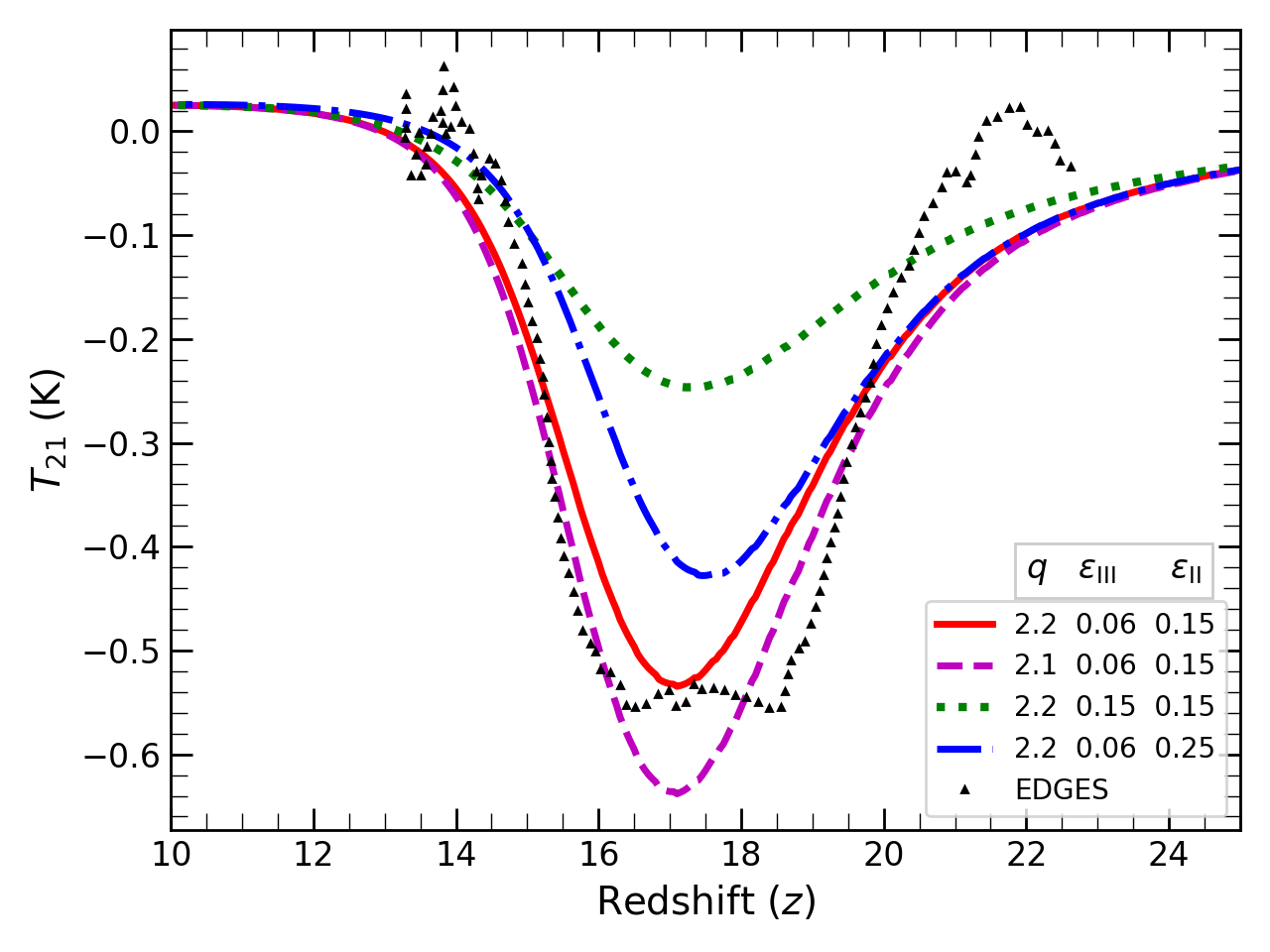}\includegraphics[width=\columnwidth]{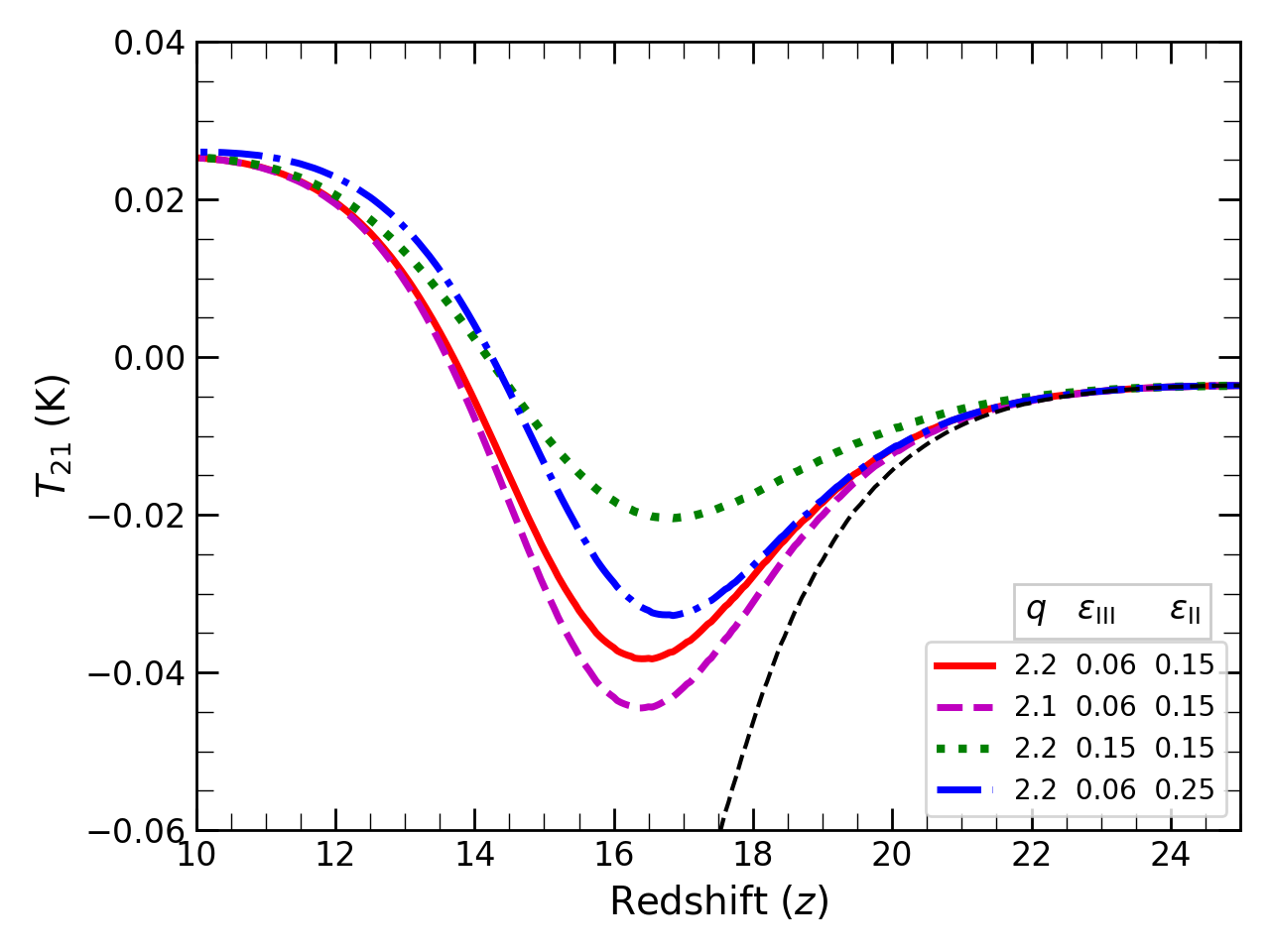}}
    \caption{{\bf Left panel:} The impact of cosmic ray heating on $T_g$, $T_s$ and resulting $T_{21}$ (upper to lower) are shown for different parameters describing the efficiencies of cosmic ray heating. The inset values represent the spectral index of cosmic ray spectra, $q$, the efficiencies of heating due to cosmic rays generated from Pop~III and Pop~II stars, $\epsilon_{\rm III}$ and $\epsilon_{\rm II}$ respectively. {We keep the other parameter, $Q_{\rm CR,II}=\epsilon_{\rm II}$ for this figure}. In this panel the evolution of temperatures are done in the extra cooling scenario due to dark matter-baryon interaction in light of EDGES observation. The black dotted curves in the upper and middle panel represent the CMBR temperature, $T_{\gamma}$, whereas the black dashed curves are the temperatures where no heating source is present.  {\bf Right panel:} Same as left panel but in absence of any dark matter-baryon interaction.}
    \label{fig:temp}
\end{figure*}   

In this section, we show the changes in gas temperature, spin temperature and resulting 21-cm brightness temperature by varying our model parameters such as $q$, $\epsilon_{\rm III}$, and $\epsilon_{\rm II}$ that govern the energy deposition by cosmic ray particles into the IGM from both Pop~III and Pop~II stars. {Due to the fact that the $\epsilon_{\rm II}$ and $Q_{\rm CR, II}$ alter the temperatures similarly, we are keeping $Q_{\rm CR, II}$ as same as $\epsilon_{\rm II}$. Further note that there are few parameters of our model like IMF of Pop~III stars, supernova energetic etc that are ill constrained and can alter our results. However, all these uncertainties affect similar ways in the final temperature history of IGM and thus variation of the efficiency parameter $\epsilon_{\rm III}$ can encompass variation in any of the above mentioned parameters.} The left panels of Fig.~\ref{fig:temp} show results in presence of the dark matter-baryon interaction with dark matter mass $m_{\chi}=0.1$ GeV and the interaction cross-section $\sigma_{45}=2$. The right panel shows results in the standard scenario without the dark matter-baryon interaction. In the left top, middle and bottom panels of Fig.~\ref{fig:temp}, $T_g$, $T_s$ and $T_{21}$ are plotted respectively with the model parameters shown in the legends. We have also shown our results for the default parameters discussed above by red solid curves.
Further in the top and middle panels, the black dotted curves represent the CMBR temperature. The black dashed curve in the top panel is the gas temperature where no cosmic ray heating is considered and the corresponding spin temperature is shown in the middle panel by black dashed curve for reference. 

The variation w.r.t the cosmic ray spectral index $q$ is shown by magenta lines where we consider a shallower slope of $q=2.1$. Given the same energy input, the number of cosmic ray particles below $30$ MeV decreases for a smaller $q$. These are the particles that take part in the collisional heating and hence the flatter slope makes the cosmic ray heating less efficient as can be seen from the figure. This further increases the absorption depth in 21-cm signal as can be seen from bottom panel.

As already discussed in the previous sections, the cosmic ray heating is directly proportional to the efficiency parameters such as, $\epsilon_{\rm III}$ and $\epsilon_{\rm II}$.
For example, it can be followed from green dotted curve that if we increase the efficiency of cosmic rays originating from Pop~III stars, $\epsilon_{\rm III}$ from $6\%$ to $15\%$, the dominance of Pop~III stars increases significantly and the IGM temperature increases more faster compared to our default model. This leads to an increase in the hydrogen spin temperature resulting in a much shallower 21-cm absorption signal of $\sim -0.2$~{K}. Similarly, if we increase the cosmic ray efficiency coming from Pop~II stars, i.e from $\epsilon_{\rm II} = 15 \%$ to $\epsilon_{\rm II} = 25 \%$, the contribution to the heating by cosmic rays from Pop~II stars increases making $T_g$ to rise faster from $z \sim 17$. This in turn increases $T_s$ which leads to a lesser depth of $\sim -0.4$~{K} $T_{21}$ signal. It can be seen from blue dash dotted curves in the left panel of Fig.~\ref{fig:temp}. Thus it is clear that any higher efficiency of cosmic ray heating would be ruled out by the EDGES absorption signal. Therefore any 21-cm signal from redshift $z \sim 14-20$ can be used to constrain early star formation, and the corresponding cosmic ray heating efficiencies.

In the right panels of Fig.~\ref{fig:temp}, we show our results without considering any dark matter-baryon interaction that is neglecting the Eq.~\ref{eq:dQ_b}. All other parameters are the same as in the left panels of the figure. As expected, in general the IGM temperature is much higher having a minimum value of $\sim 10$~K by $z \sim 17$ for our default parameter set (solid red curves). This resulted in a spin temperature of $\sim 20$~K due to the Ly-$\alpha$ coupling. Afterwards, the IGM temperature and hence $T_s$ increases due to the rapid cosmic ray heating. Finally $T_s$ crosses the CMBR temperature by $z \sim 14$ which leads to a weaker absorption profile of maximum {$\sim -0.04$~K} as can be seen from bottom right panel. The exact position of the absorption maxima depends on the interplay of Ly-$\alpha$ coupling and the onset of cosmic ray heating. For example more efficient cosmic ray heating as shown by the green dotted line resulted in an absorption depth of {$\sim -0.02$~K} at an earlier redshift. Therefore, we conclude that the cosmic ray heating is likely to play an important role in shaping the global 21-cm signal. In particular, it is likely to reduce the depth of any absorption signal if at all present.
In passing we note that, the Pop~II stars are likely to heat the IGM by magnetosonic heating which is sensitive to the IGM magnetic field. Thus the global 21-cm signal can, in principle, be used to constrain the magnetic field during the cosmic dawn \citep{Minoda19, Bera_2020}.


{
\subsection{Comparison with X-ray heating}
\begin{figure}
    \centering
	\includegraphics[width=\columnwidth]{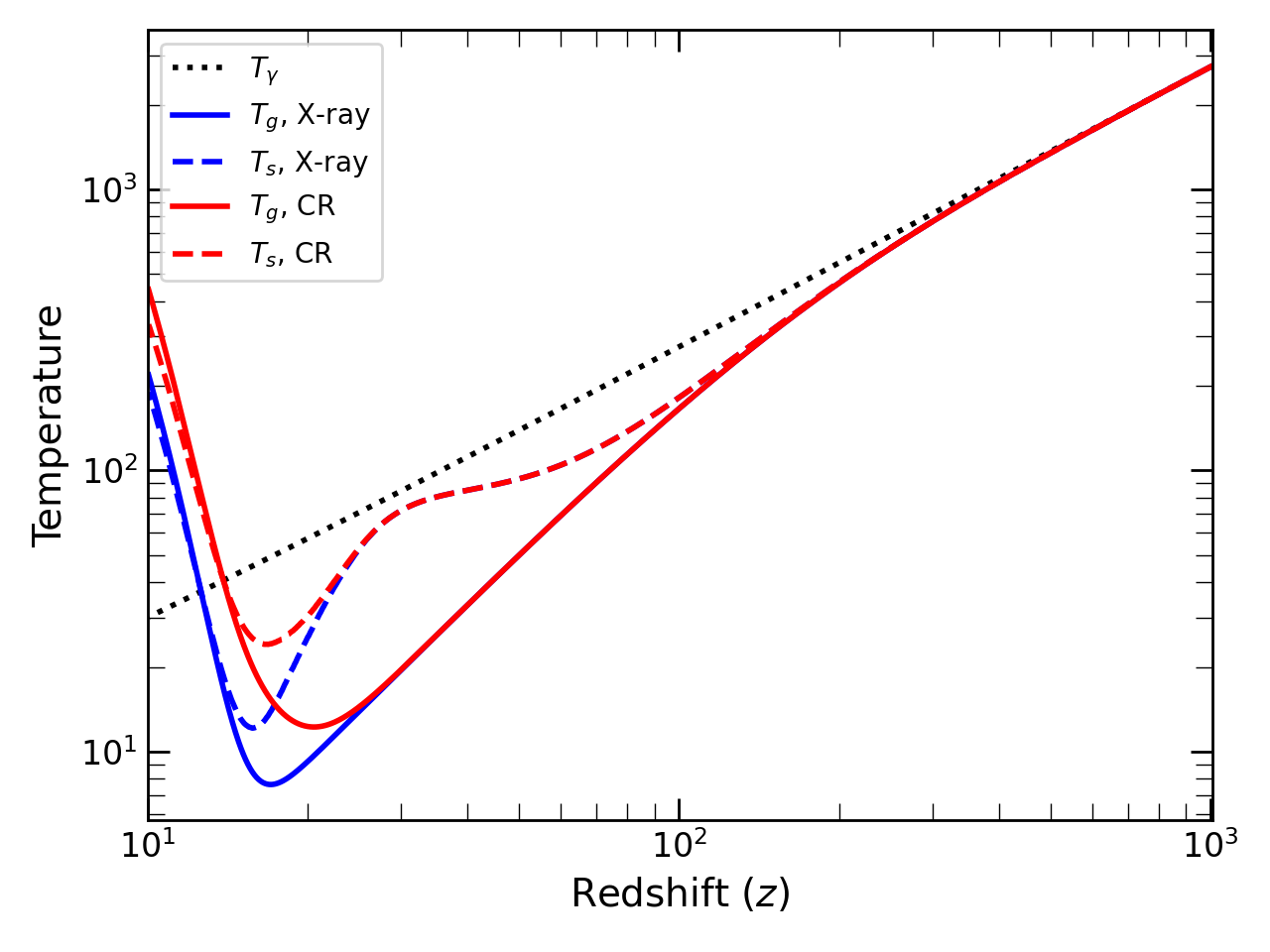}
	\includegraphics[width=\columnwidth]{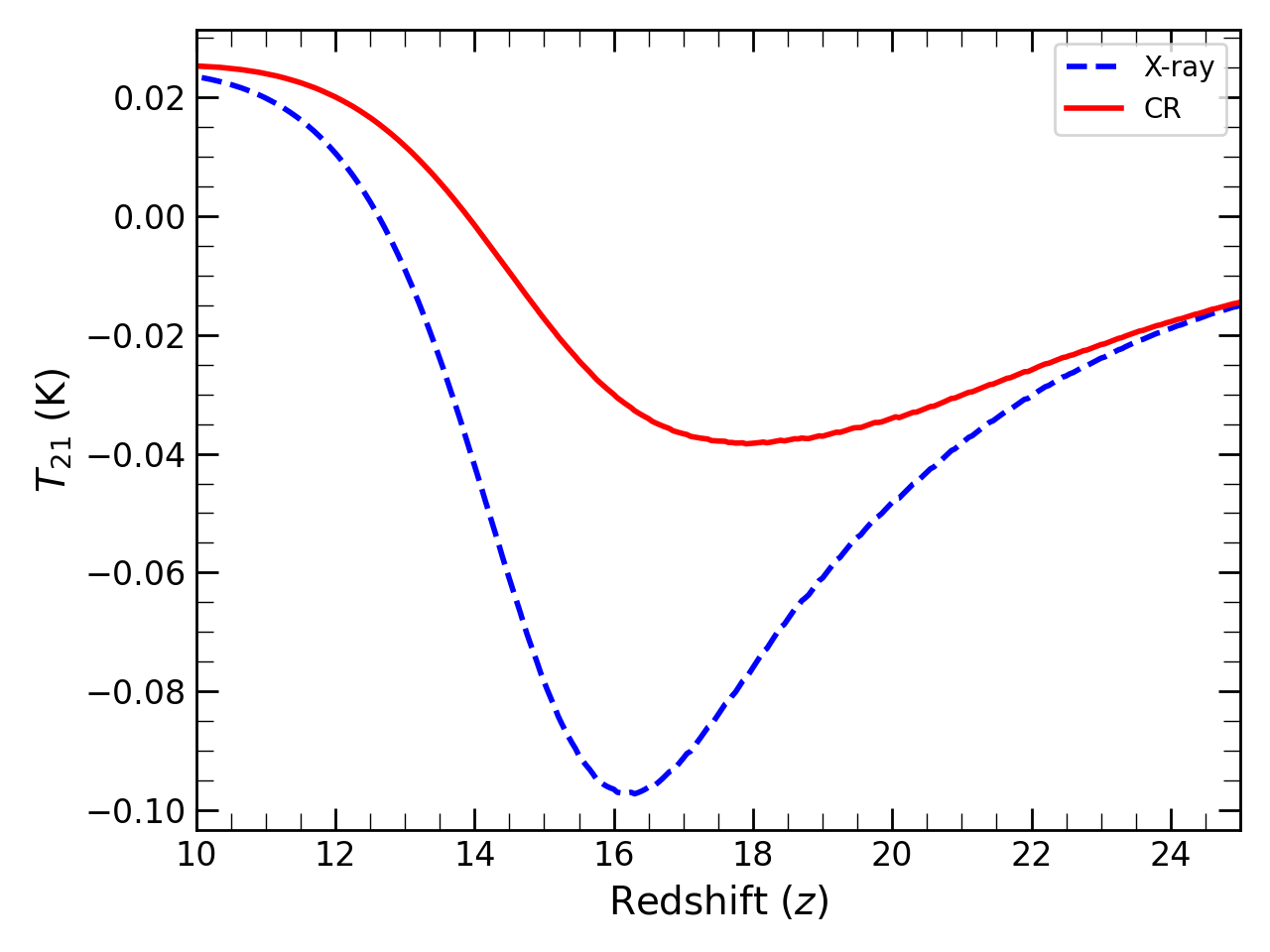}
    \caption{{The temperature evolution shown in red is due to the presence of cosmic ray heating for the parameters $\epsilon_{\rm III}=0.1$, $\epsilon_{\rm II}$ and/or $Q_{\rm CR,II}=0.15$, while the temperatures shown in blue are due the X-ray heating for the parameters $f_{\rm Xh,III}=0.001$, $f_{\rm Xh,II}=0.2$. The resulting brightness temperatures, $T_{21}$ are shown in the lower panel due to X-ray heating (blue dashed) and cosmic ray heating (red solid).}}
    \label{fig:Temp_CR_Xray}
\end{figure}
There are other processes that are likely to be responsible for heating of the inter-galactic medium like X-ray heating \citep{pritchard2007, baek09, ghara15, fialkov14b}, Ly-$\alpha$ \citep{Chuzhoy2006, ghara2020, mittal21}and CMB \citep{Venumadhav2018} etc. At present there is no evidence showing the domination of any heating mechanism over others. All possible heating mechanisms have their own free parameters which are not well constrained. Thus only when more observational evidences are available, a detailed modeling of 21~cm signal including all possible sources of heating is effective and that is beyond the scope of the current work. However, to demonstrate the importance of cosmic ray heating, we compare our results due to cosmic rays heating with the X-ray heating without the presence of any exotic physics such as dark matter-baryon interaction.
In the upper panel of Fig.~\ref{fig:Temp_CR_Xray}, we have shown the comparison of gas (solid) and spin (dashed) temperatures for cosmic ray heating (red) and X-ray heating (blue) for the same star formation rate densities and the Ly-$\alpha$ coupling. In order to include the X-ray heating in our model we followed Eq.~14 given in \citet{atri2021}. In case of cosmic ray heating we have taken the efficiencies as $\epsilon_{\rm III}=0.1$ and $\epsilon_{\rm II}$ and/or $Q_{\rm CR,II}=0.15$ for this particular plot. On the other hand, for X-rays the contribution to the heating from Pop~III stars and Pop~II stars are $f_{\rm Xh,III}=0.001$ and $f_{\rm Xh,II}=0.2$ which is motivated by \citet{atri2021} and \citet{furlanetto06}. We varied the heating rate due to X-rays coming from Pop~III stars from $f_{\rm Xh,III}=0.001$ to $f_{\rm Xh,III}=0.1$ and the changes in temperatures are negligible as stated in \citet{atri2021}. It is clear from the blue and red curves that the contribution of cosmic rays and X-rays to the heating are quite similar. Although a small change in $T_s$ can make a big difference in the resulting brightness temperature that is shown in the lower panel of Fig.~\ref{fig:Temp_CR_Xray}. For the set of parameters that we used here, the cosmic ray heating is more efficient as compared to the X-ray heating for the same star formation history, though depending upon the parameters the magnitude of heating can be changed. Thus we conclude that the cosmic ray heating is comparable to X-ray heating during cosmic dawn and has similar effect to shape the global 21~cm signal.
}

\section{Summary and Conclusions}
\label{main_sec:summary}

We present a semi-analytical model of global HI 21-cm signal from cosmic dawn focusing on the heating by cosmic rays generated from the early generation of stars. In our model, we take into account both metal free Pop~III and metal enriched Pop~II stars. We consider the meta-galactic Lyman-Werner feedback on the Pop~III stars along with supernova feedback in Pop~II galaxies in order to calculate the star formations. The global HI 21-cm signal is calculated self-consistently taking into account the Ly-$\alpha$ coupling. The temperature evolution of IGM has been calculated using the cosmic ray heating along with the dark matter-baryon interaction and adiabatic cooling. We consider the ionization and collisional heating by low energy cosmic ray protons generating and escaping from Pop~III halos.
Besides accounting for the amount of cosmic ray energy in the IGM, our model captures the evolution of the cosmic ray spectrum, as they propagate through the IGM and lose their energy by ionizing and exciting neutral Hydrogen atoms, and due to the adiabatic expansion. We further consider the cosmic rays produced by the Pop~II stars which transfer energy to the inter-galactic medium through the generation of magnetosonic Alfv\'en waves.
{Here, we would like to mention that, all the quantities such as IGM kinetic/spin temperature, ionization fraction, Lyman-$\alpha$/Lyman-Werner background radiation, star formation rate etc estimated here are globally averaged quantities. In principle, one should first simulate them on 3D cubes self-consistently at different redshifts and then perform the averaging. We must note that,  although estimating the global 21-cm signal directly from the global quantities is faster, it can introduce bias in the estimated signal.}

{We find that the cosmic ray is an important source of IGM heating and shapes the global 21-cm signal during cosmic dawn which is in agreement with the existing literature.} The depth, duration and timing of the absorption signal are highly modulated by the cosmic ray heating that we consider here. In particular, the EDGES signal can be well explained by our model of cosmic ray heating along with the Ly-$\alpha$ coupling and dark matter-baryon interaction with a suitable choice of efficiency parameters.
In fact, the required efficiency parameters of cosmic ray heating are reasonably small like $\epsilon_{\rm III}=0.06$ \& $\epsilon_{\rm II}$ {and/or $Q_{\rm CR,II}=0.15$} to produce significant heating of the IGM and match the EDGES observed profile. Further, we explore the various efficiency parameters related to the cosmic ray heating and show that the brightness temperature highly depends on these parameters. In particular, highly efficient cosmic ray heating reduces $T_{21}$ by a significant amount. We also showed that the cosmic rays can highly impact the IGM temperature or in turns the 21-cm signal, in absence of any dark matter-baryon interaction, and could even potentially wash out the absorption signal during cosmic dawn. 
Thus cosmic rays need to be considered as a potential source of IGM heating along with widely explored sources of heating such as through soft X-rays during the cosmic dawn. 
We further argued that, since the cosmic 21-cm signal can be highly modulated by the heating due to cosmic rays produced by the early generation of stars, accurately determined 21-cm signal by experiments such the EDGES  \citep[Experiment to Detect the Global Epoch of Reionization Signature,][]{EDGES18}, SARAS  \citep[Shaped Antenna measurement of the background RAdio Spectrum,][]{raghunathan21}, LEDA  \citep[Large-aperture Experiment to Detect the Dark Ages,][]{price18}, REACH  \citep[Radio Experiment for the Analysis of Cosmic Hydrogen,][]{acedo19}, SKA (Square Kilometer Arrays), HERA (Hydrogen Epoch of Reionization Array)  etc.  could be used to probe the early cosmic ray heating and constrain the nature of these early generation of stars.  More detailed studies such as comparing the efficiency of cosmic ray heating with other potential sources like soft X-rays and prospects of constraining early cosmic ray heating using cosmic HI 21-cm signal are deferred for future works.

\section*{Acknowledgements}

AB acknowledges financial support from UGC, Govt. of India. SS and KKD acknowledge financial support from BRNS through a project grant (sanction no: 57/14/10/2019-BRNS).
We thank Erik Zackrisson for useful discussion. KKD acknowledges financial support from SERB-DST (Govt. of India) through a project under MATRICS scheme (MTR/2021/000384). SS thanks Presidency University for the support through FRPDF grant.

\section*{DATA AVAILABILITY}
The data underlying this work will be shared upon reasonable request to the corresponding author.




\bibliographystyle{mnras}
\bibliography{refs} 





\bsp	
\label{lastpage}
\end{document}